\pdfoutput=1
\documentclass[aps,prl,twocolumn,superscriptaddress,showpacs]{revtex4-1}

\usepackage{graphicx}
\usepackage{afterpage}

\begin{document}


\title{Magnetic, thermodynamic, and electrical transport properties of
the noncentrosymmetric B20 germanides MnGe and CoGe}



\author{J. F. DiTusa}
\email[]{ditusa@phys.lsu.edu}
\affiliation{Department of Physics and Astronomy, Louisiana State
University, Baton Rouge, Louisiana 70803, USA}

\author{S. B. Zhang}
\affiliation{International Center for Materials Nanoarchitectonics
(MANA), National Institute for Materials Science, Tsukuba, Ibaraki
305-0044 Japan}

\author{K. Yamaura}
\affiliation{Superconducting Properties Unit, National Institute for
Materials Science 1-1 Namiki, Tsukuba, 305-0044 Ibaraki, Japan}

\author{Y. Xiong}
\affiliation{Department of Physics and Astronomy, Louisiana State
University, Baton Rouge, Louisiana 70803, USA}

\author{J. C. Prestigiacomo}
\affiliation{Department of Physics and Astronomy, Louisiana State
University, Baton Rouge, Louisiana 70803, USA}

\author{B. W. Fulfer}
\affiliation{Department of Chemistry, Louisiana State University,
Baton Rouge, Louisiana 70803, USA}

\author{P. W. Adams}
\affiliation{Department of Physics and Astronomy, Louisiana State
University, Baton Rouge, Louisiana 70803, USA}

\author{M. I. Brickson}
\affiliation{Goshen College, Goshen, IN 46526, USA}

\author{D. A. Browne}
\affiliation{Department of Physics and Astronomy, Louisiana State
University, Baton Rouge, Louisiana 70803, USA}

\author{C. Capan}
\affiliation{Department of Physics and Astronomy, Washington State
University, Tri-City Campus, Pullman, Washington 99164, USA}

\author{Z. Fisk}
\affiliation{Department of Physics and Astronomy, University of
California, Irvine, Irvine, California 92697, USA}

\author{Julia Y. Chan}
\affiliation{Department of Chemistry, University of Texas at Dallas,
Richardson, TX 75080, USA}


\date{\today}

\begin{abstract}
We present magnetization, specific heat, resistivity, and Hall effect
measurements on the cubic B20 phase of MnGe and CoGe and compare to
measurements of isostructural FeGe and electronic structure
calculations. In MnGe, we observe a transition to a magnetic state at
$T_c=275$ K as identified by a sharp peak in the ac magnetic
susceptibility, as well as second phase transition at lower
temperature that becomes apparent only at finite magnetic field. We
discover two phase transitions in the specific heat at temperatures
much below the Curie temperature one of which we associate with
changes to the magnetic structure. A magnetic field reduces the
temperature of this transition which corresponds closely to the
sharp peak observed in the ac susceptibility at fields above 5
kOe. The second of these transitions is not affected by the
application of field and has no signature in the magnetic properties
or our crystal structure parameters. Transport measurements indicate
that MnGe is metal with a negative magnetoresistance similar to that
seen in isostructural FeGe and MnSi. Hall effect measurements reveal a
carrier concentration of about 0.5 carriers per formula unit also
similar to that found in FeGe and MnSi. CoGe is shown to be a low
carrier density metal with a very small, nearly temperature
independent diamagnetic susceptibility.
\end{abstract}


\maketitle

\section{I. Introduction}
\begin{figure}[htb]
  \includegraphics[angle=0,width=3.0in,bb=100 100 620
  520,clip]{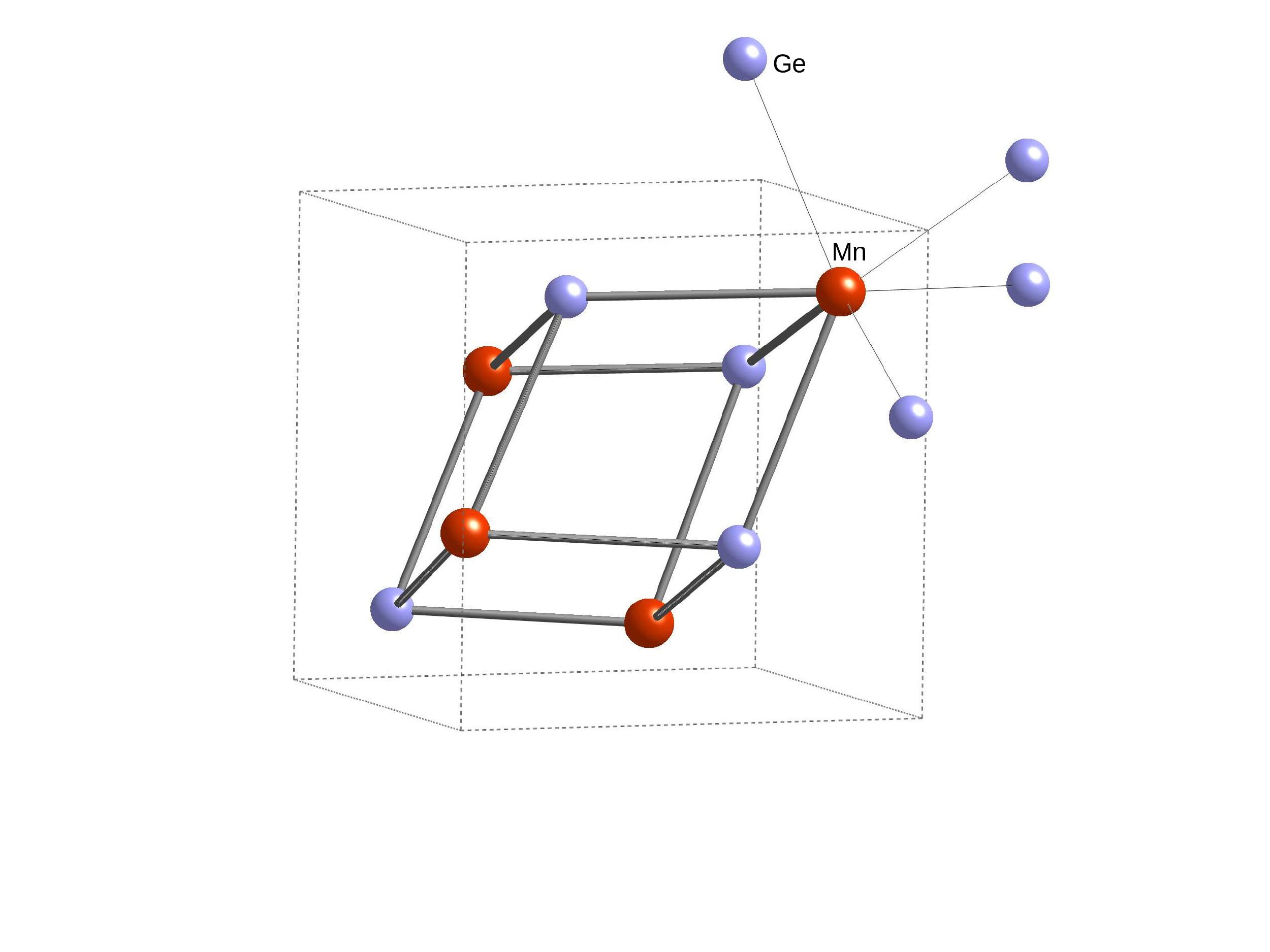}%
  \caption{\label{fig:struc} Crystal Structure. Schematic of the cubic
B20 crystal structure adopted by a number of the transition metal
({\it TM}) germanide and silicide compounds including MnGe, MnSi, FeGe,
FeSi, CoGe and CoSi. The unit cell is indicated by the dotted
lines. One of the important and interesting features of this structure
evident in this figure is its lack of inversion symmetry. {\it TM} atoms
occupy 4 sites that form a tetrahedron aligned along the (111)
direction along with the 4 nonmetals positioned on a tetrahedron
inverted relative to the {\it TM} atoms. }
\end{figure}

The silicides of Cr, Mn, Fe, and Co all which form in the B20 crystal
structure type shown in Fig.~\ref{fig:struc}, notable because of its
lack of inversion symmetry, have been investigated for over 40 years
yet continue to yield fascinating discoveries\cite{wernick}. Although
CrSi and CoSi appear to be simple paramagnetic (PM)
metals\cite{wernick}, MnSi is helimagnetic (HM) below 30
K\cite{ishikawa2,ishikawa} and FeSi is a small band gap insulator with
unusual temperature, $T$, dependent
properties\cite{wernick,jaccarino,aeppli,schlesinger,delaire}. Chemical
substitutions among these materials, such as Fe$_{1-x}$Mn$_x$Si and
Fe$_{1-y}$Co$_y$Si, also display helimagnetism over wide regions of
$x$ and $y$ as well as interesting behavior near the
insulator-to-metal transitions for $x, y \sim
0.01$\cite{beille,manyala1,manyala2,manyala3}. The low symmetry of the
B20 crystal structure gives rise to large Dzyaloshinskii-Moriya
interactions causing long-period HM, rather than ferromagnetic (FM),
ground states\cite{nakanishi,lebech}.

Investigations into the properties of MnSi have shown that the Curie
temperature, $T_c$, is readily reduced by moderate pressure, $P$, with
$T_c$ approaching zero near 14 kbar\cite{pfleiderer1}. Surprisingly,
instead of accessing a quantum critical point with $T_c$ going to
zero, this system avoids criticality by undergoing a transition to an
unusual state thought to be a crystal of topologically stable knots of
the spin structure known as a Skyrmion
lattice\cite{pfleiderer2,pfleiderer3,muhlbauer,neubauer1,munzer,yu1,yu2}. There
is also evidence for this unusual magnetic phase at ambient $P$ over a
small range of magnetic field, $H$ and $T$ near $T_c$, a region
previously labeled as the $A$-phase\cite{muhlbauer}. Evidence for a
Skyrmion lattice in thin samples Fe$_{1-y}$Co$_y$Si and in FeGe
exposed to small fields was discovered in Lorentz force microscopy
images which observe a
circulating magnetic moment on the length scale of the helimagnetic
periods that exist at $H=0$\cite{yu1,yu2}.

\begin{figure}[htb]
  \includegraphics[angle=90,width=3.2in,bb=0 0 440
  360,clip]{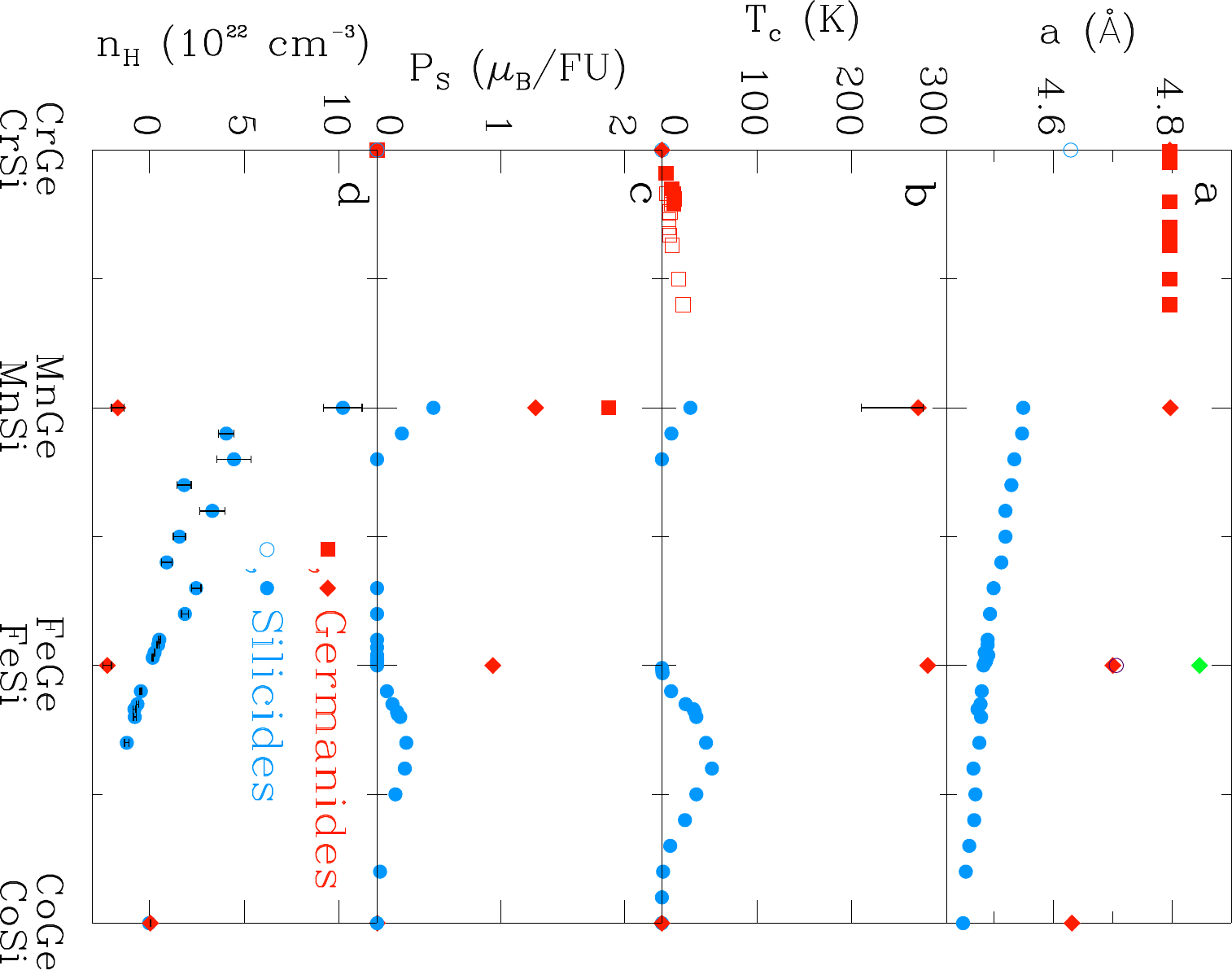}%
  \caption{\label{fig:tccomp} Phase diagram of the monogermanides and
monosilicides having the B20 crystal structure. a) lattice constant,
$a$. Data for CrSi, open circle, from Ref.\cite{vandermarel}. Green
diamond is RuGe and violet circle is RuSi taken from
Ref.~\protect{\cite{hohl}}. b) Curie temperature, $T_c$, closed
symbols, and spin glass transition temperature, open symbols. Error
bar drawn for MnGe demonstrates the the difference in our data between
$T_c$ indicated in the low-field magnetic susceptibility and that
obtained from a mean-field analysis of the high field
magnetization. c) Saturation magnetization, $P_S$ as determined from
the magnetization at high magnetic field. Filled square for MnGe
indicates the value found in Ref.~\cite{kanazawa} at fields (150 kOe)
not accessed in our experiments. d) carrier concentration, $n_{H}$,
determined from the ordinary Hall constant at 5 K for several
germanide and silicide compounds identified on the $x$-axis. Data for
Mn through Co silicide taken from Ref.~\cite{manyala2}. Data for
Cr$_{1-x}$Mn$_x$Ge for $x\le 0.6$ in frames a, b, and c, indicated
with filled and open squares, taken from
Ref.~\protect{\cite{sato1,sato3}} with permission from the publisher.
}
\end{figure}

In addition to these interesting magnetic properties, these materials
may be important from a spintronics
viewpoint\cite{manyala2}. Silicides are intrinsically compatible with
silicon technologies and there are both insulating and metallic
magnetic states with high carrier spin polarizations in this series.
In addition, chemical substitutions between the monosilicides form
easily and show little proclivity toward nucleating second phases.
Thus, they may prove ideal for spin injection into silicon
devices\cite{manyala2}.  However, as the $T_c$'s only reach 65 K for
Fe$_{0.6}$Co$_{0.4}$Si, see Fig.~\ref{fig:tccomp}, they are not likely
to be useful for most applications\cite{wernick,manyala1}.

Given all of this interest, it is natural to ask if there are other
isostructural materials that might also yield interesting magnetic
ground states. Although CrGe is the only 3d {\it TM} germanide to have an
equilibrium B20 crystal structure\cite{sato1}, FeGe crystallizes
readily in this structure when the growth conditions are carefully
controlled\cite{richardson}. MnGe and CoGe are also known to form in
this structure, but only under conditions of high pressure and
temperature\cite{takizawa}. CrGe, much like chromium silicide, forms a
nonmagnetic metal\cite{sato1,sato2}. In contrast, FeGe has been the
subject of much recent interest as it is very different from
insulating FeSi displaying a metallic and HM ground state with a
period of 70 nm and a $T_c=280$ K\cite{lundgren,lebech,yeo}. The
$A$-phase of FeGe has recently been explored by Wilhelm {\it et al.}
where they find a rich set of transitions that they suggest are due to
symmetry changes stemming from solitonic inter-core interactions and
the onset of chiral modulations\cite{wilhelm1,wilhelm2}. Electronic
structure calculations are interesting in that LDA, which correctly
predicts a small band-gap insulating state for FeSi, predicts an even
smaller band gap for the paramagnetic state for
FeGe\cite{anisimov}. The metallic and magnetic ground state can be
achieved by adding an on-site Coulomb repulsion ($U$)\cite{anisimov}.
Thus, electronic structure calculations confirm that FeGe is a
metallic magnet and reveal that it has a nearly complete conduction
electron spin polarization at low $T$. That FeGe has a $T_c$
approaching room temperature, and hosts a Skyrmion lattice phase for
$T$ just below $T_c$\cite{yu2}, suggests that there may be related
materials with $T_c$s large enough to be considered for applications.

Here we investigate the synthesis, magnetic, thermodynamic, and charge
carrier transport properties of MnGe and CoGe to compare their
properties to those found in the isostructural silicides with the goal
of exploring the complex magnetic states of MnGe. We have included
comparisons to calculations of their electronic structure and
measurements of FeGe since the magnetic behavior of this compound is
thought to be well
established\cite{lebech,wilhelm1,wilhelm2}. Previous investigations of
MnGe reveal that it is very likely HM with a period that increases
from 4 to 8 nm from 30 to 150 K with evidence that it hosts a Skyrmion
lattice over a wide range of $T$ and
$H$\cite{kanazawa,makarova,kanazawa2}. Although CoGe has properties
similar to its periodic table neighbor CoSi\cite{kanazawa3}, being a
diamagnetic metal with a low carrier density, our data indicate that
MnGe may be more complex than either MnSi or FeGe. Our ac and dc
susceptibility data at low fields indicate a magnetic transition at
275 K while our magnetization data reveal a large saturated magnetic
moment of $>1.3$ $\mu_B/$FU. Kanazawa {\it et al.}\cite{kanazawa} have
previously demonstrated that at low $T$ the magnetization saturates
near 2 $\mu_B$/FU at fields above those probed here. Several phase
transitions are evidenced by sharp peaks in both our ac susceptibility
and specific heat data between 70 and 165 K that have not been
previously observed. We suggest that these may be associated with the
complex phase diagram expected in systems hosting Skyrmion lattices as
has been observed very near $T_C$ in MnSi\cite{bauer} and
FeGe\cite{wilhelm1,wilhelm2}.  If this is true, the large $T$ and $H$
ranges where the Skyrmion lattice state appears to be stable in
MnGe\cite{kanazawa2} implies that it may be an ideal compound to
explore these topologically interesting magnetic phases. A second
phase transition at 120 K is also apparent in the specific heat that
has no counterpart in the magnetic measurements and whose origin is
still unknown. The transport properties of MnGe are very much like
those of FeGe, but with a larger field scale consistent with the
larger saturation field.

Our electronic structure calculations  for MnGe are largely consistent
with the  above description with a ferromagnetic ground state and a
large  carrier polarization  energetically  favored. Interestingly,  a
half metallic state appears to be stable at a somewhat smaller lattice
constant\cite{robler}. We  present these results,  along with the  calculated Fermi
surfaces,  and compare  the equilibrium  lattice constant  and reduced
lattice constant solutions. Our results for CoGe indicate a very small
carrier density metal with a correspondingly small Fermi surface (FS).

\section{II. Experimental Details}

Polycrystalline samples of MnGe and CoGe were prepared from high
purity starting materials purchased from Wako Pure Chemical
Industries, Ltd. and Rare Metallic Co. Ltd. The raw materials were
checked for purity and the Ge powder was reduced in a 5\% H$_2$ in Ar
flow at 750 $^o$C to remove any remaining oxide. These starting
materials were ground into a fine powder and sealed in Ta/BN capsules.
Samples were heated to 1200 $^o$C for CoGe and 1300 $^o$C for MnGe for
a period of 1.5 h at a pressure of 6 GPa. Samples with excess Ge,
CoGe$_{1.1}$ and MnGe$_{1.2}$ were grown under the same conditions in
order to check the effect of any impurity phases on our results. X-ray
powder diffraction data were obtained at room $T$ using a Philips
X'pert X-ray diffractometer with Cu$K_\alpha$ radiation and confirmed
using a Bruker Advance D8 powder diffractometer equipped with a
focusing Ge(111) incident beam monochromator (Cu$K\alpha_1$
radiation). CoGe samples showed no indication of any second phases and
were determined to have the B20 crystal structure with a lattice
constant, $a$, of 4.631 $\AA$ while CoGe$_{1.1}$ was also determined
to have a B20 crystal structure with a larger lattice constant of
4.639 $\AA$. MnGe samples displayed the same B20 crystal structure
with $a=4.797(4) \AA$. However our powder diffraction patterns
revealed the presence of a small amount second phase thought to be the
high temperature phase Mn$_2$Ge in our MnGe sample ($\sim 5$\%) with a
larger fraction evident in our MnGe$_{1.2}$ sample ($\sim 20$\%). In
addition, a small single crystal of MnGe was extracted from our sample
and mounted on a glass fiber in a Nonius Kappa CCD diffractometer (Mo
$K\alpha$, $\lambda=0.71073 \AA$) for single crystal X-ray diffraction
measurements. Refinement of the crystal structure using 107
reflections confirmed the B20 structure with a lattice constant of
4.797(4) $\AA$. In the B20 structure (Wyckoff \#198, P213), the atoms
are at ($u$,$u$,$u$), ($1/2-u$,$1-u$,$1/2+u$), ($1-u$,$1/2+u$,$1/2-u$)
and ($1/2+u$, $1/2-u$, $1-u$). Our refinement determined $u$ to be
0.1377(2) for Mn and 0.84388(15) for Ge. Data were collected between
95 and 300 K that showed a thermal contraction of the lattice constant
by $\sim 1$\% with cooling. The subtle crystal structure symmetry
change at 170 K suggested in Ref.~\cite{makarova} was not observed
although the consequences of this purported structural transition are
near the limits of our ability to detect.

Single crystals of FeGe were grown by standard vapor transport
techniques previously described in Ref.~\cite{richardson} and
\cite{yeo}. A stoichiometric mixture of high purity elements
were arc melted and then placed in an evacuated quartz tube along with
the transport agent, iodine. These were heated in a two zone furnace
for 1 week. Powder X-ray diffraction was employed to check the phase
purity.

Magnetic susceptibility, $\chi$, and magnetization, $M$, measurements
were performed in a Quantum Design (QD) MPMSXL SQUID magnetometer in a
50 kOe superconducting magnet from 2.5 to 400 K. Both dc and ac
susceptibility measurements were performed with the ac $\chi$ taken
with an excitation fields of 1 Oe at a frequency of 30 Hz.  Specific
heat measurements were performed in a QD PPMS using a standard
semi-adiabatic heat pulse technique from 2 to 300 K. The specific heat
of MnGe was performed in fields of 0, 10, and 30 kOe provided by a
superconducting magnet. Data presented here have been corrected by
carefully subtracting the contribution from the measurement addenda.
The electrical conductivity and Hall effect measurements on
polycrystalline samples of MnGe and CoGe and single crystalline samples
of FeGe were performed on rectangular shaped samples polished with
emery paper. Thin Pt wires were attached to four Epotek silver epoxy
contacts with an average spacing between the voltage probes of 0.5
mm. Samples had an average cross sectional area of 1.0 x 0.1
mm$^2$. The resistivity, magnetoresistance and Hall effect
measurements were performed at 17 Hz using standard lock-in techniques
in a gas flow cryostat and a 50 kOe superconducting magnet. Hall effect
measurements were corrected for any misalignment of the leads by
symmetrizing the data collected at positive and negative fields.

\section{III. Experimental Results}
\subsection{A. Magnetic Susceptibility and Magnetization}
\begin{figure}[htb]
  \includegraphics[angle=90,width=2.8in,bb=0 0 320
  420,clip]{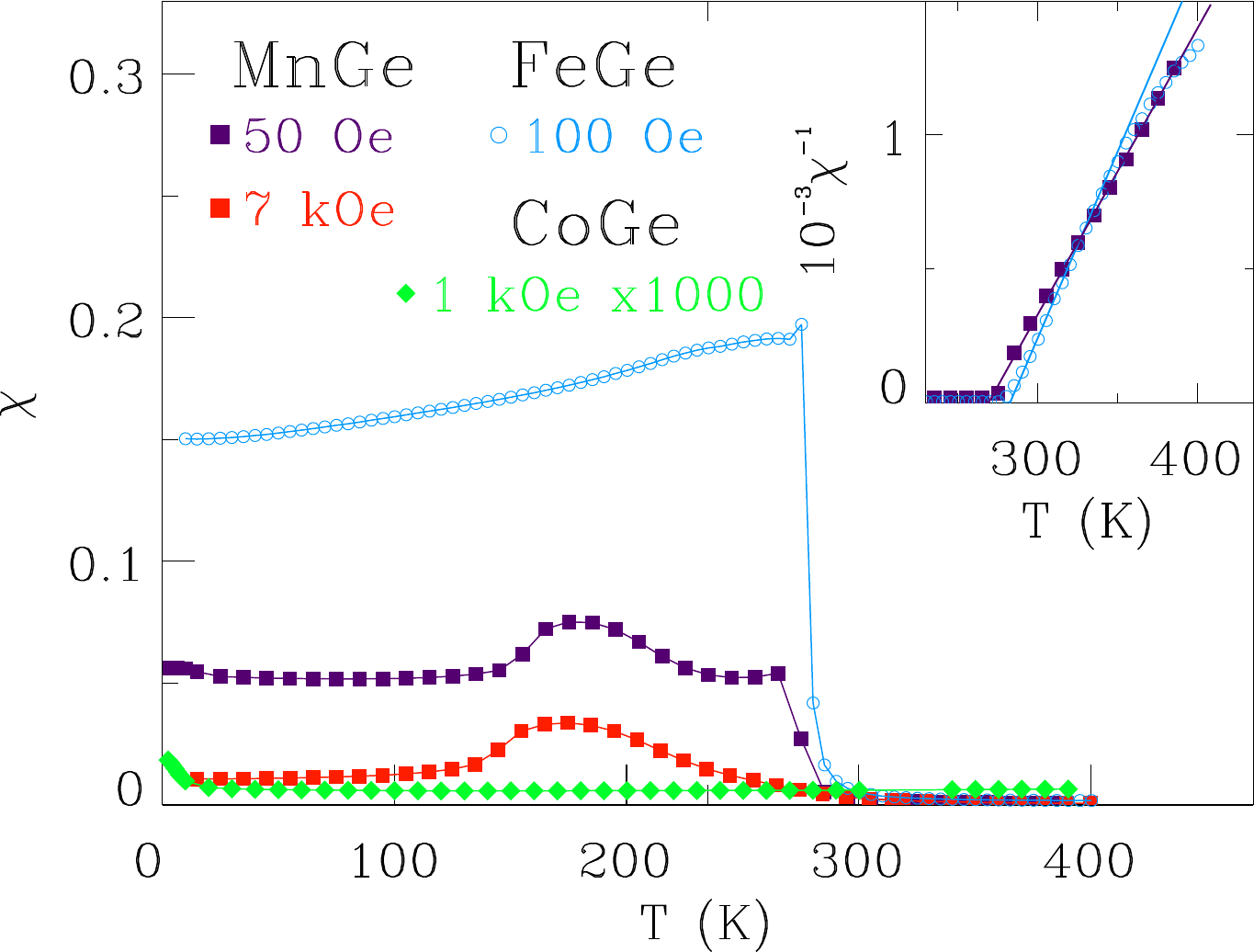}%
  \caption{\label{fig:chifig} Magnetic Susceptibility. dc Magnetic
susceptibility, $\chi$, vs.\ temperature, $T$, for FeGe, MnGe and CoGe
at magnetic fields indicated in the figure. Note that the CoGe data
has been multiplied by a factor of 1000 for clarity. Inset:
$\chi^{-1}$ vs.\ $T$ for FeGe and MnGe. Lines represent fits to
Curie-Weiss behavior and correspond to Weiss temperatures of 284 and
270 K and magnetic moments of 1.0 and 1.35 $\mu_B$ per formula unit for
FeGe and MnGe, respectively.}
\end{figure}

The dc $\chi$ of polycrystalline samples of both MnGe and CoGe are
displayed in Fig.~\ref{fig:chifig} along with $\chi$ of a single
crystal of FeGe.  The magnetic susceptibility of FeGe is
quantitatively similar to that of previous
measurements\cite{yeo}. Measurements performed on the single crystal
of MnGe separated from the same melt (without orientation) were nearly
identical with those shown here for the polycrystalline sample. The
general features of the magnetic susceptibility of MnGe that we
observe are similar to that measured previously\cite{kanazawa},
however, here we find a larger $\chi$ at low fields and a magnetic
transition is observed near 275 K. We present $\chi$ of MnGe at 50 Oe
and 7 kOe to display the variability with moderate fields that we
observe. For temperatures above 290 K, Curie-Weiss behavior is
apparent for both FeGe and MnGe as demonstrated in the inset.  Here
the lines are fits of this form with a effective moment of 1.0 and
1.35 $\mu_B$ per formula unit (FU) and a Weiss temperature of 284 and
270 K for FeGe and MnGe respectively. We have checked that these
values are independent of $H$ for $H\le 50$ kOe for MnGe. The magnetic
susceptibility of CoGe is paramagnetic, but much smaller, requiring an
amplification by a factor of 1000 to make it visible on the scale of
Fig.~\ref{fig:chifig}. A small upturn is seen below 20 K which is most
likely due to a small density of paramagnetic impurities.

\begin{figure}[htb]
  \includegraphics[angle=90,width=2.8in,bb=0 0 450
  420,clip]{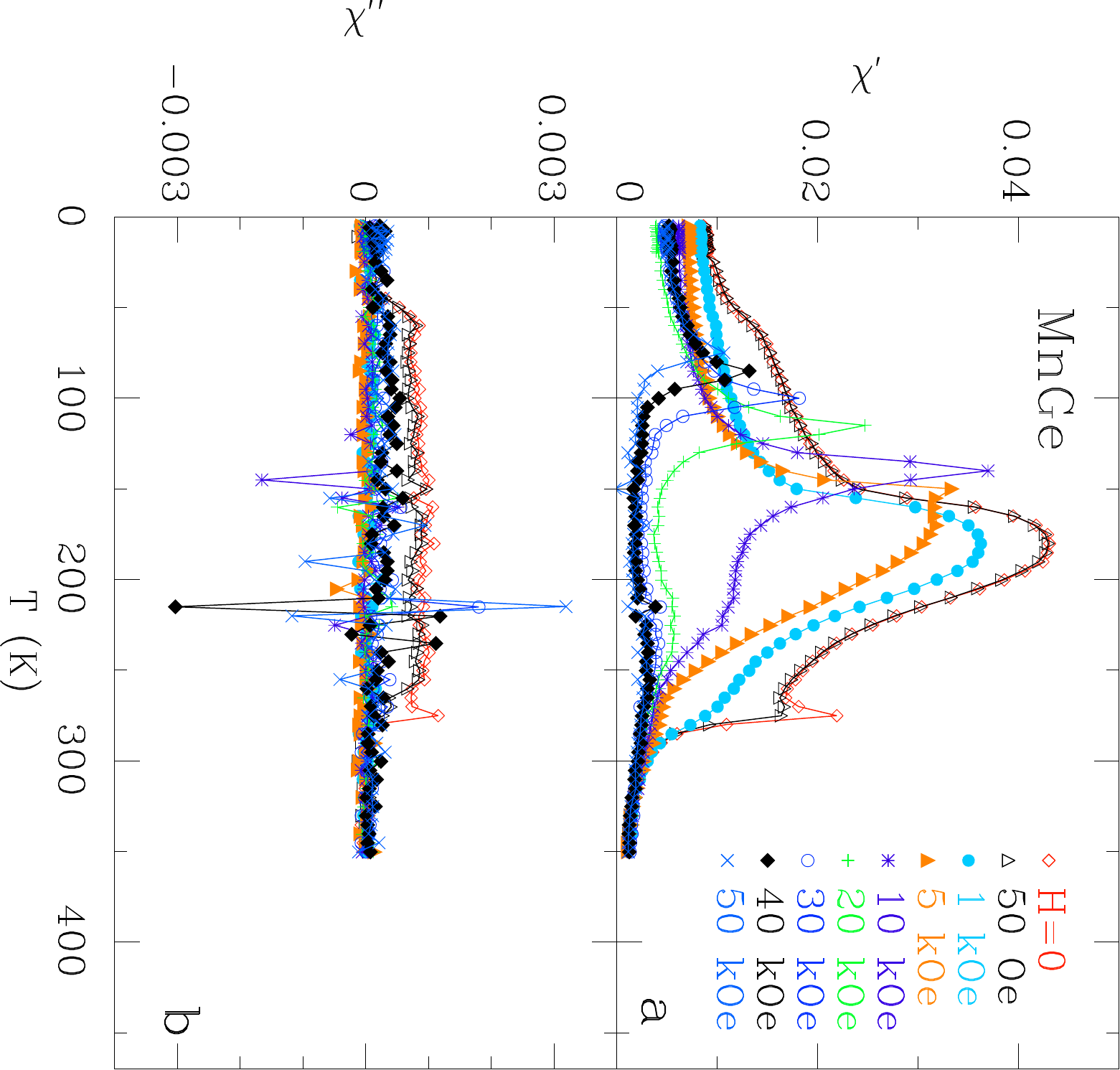}%
  \caption{\label{fig:chiacfig} ac Magnetic Susceptibility. a) Real, $\chi'$, and
b) imaginary, $\chi''$, parts of the ac magnetic susceptibility vs.\ $T$
for MnGe at magnetic fields indicated in the figure. Lines connect the
data points for display purposes.  }
\end{figure}

To explore more fully the subtle features apparent in our $\chi$
measurements of MnGe, we have measured the ac susceptibility over a
wide range of $T$ and $H$. The real part of the ac magnetic
susceptibility, $\chi'$, of MnGe is displayed in
Fig.~\ref{fig:chiacfig}a and above 300 K yields a Weiss $T$ and Curie
constant within error of our dc results.  We observe a sharp peak in
the low field $\chi'$ at 275 K in agreement with the peak we observed
in the low field dc $\chi$ measurements. Below 250 K a large increase
in $\chi$ and $\chi'$ is followed by a broad maximum near 180 K. This
broad peak is suppressed with field while a sharp peak in $\chi'$ 
becomes apparent at lower $T$ for $H\ge 5$ kOe. These
sharp features are indicative of a phase transition which moves to
lower $T$ with increasing $H$. The corresponding imaginary part of the
ac susceptibility, $\chi''$ shown in frame b of
Fig.~\ref{fig:chiacfig} displays only a few features that can be
considered as signal above the background. In particular, $\chi''$ at
fields 0 and 50 Oe is enhanced between 50 and 275 K which corresponds
well to the region between the low temperature shoulder and the sharp
peak at 275 K in $\chi'$.  In addition, resonance features are
apparent near 210 K for $H \ge 20$ kOe.

\begin{figure}[bht]
  \includegraphics[angle=90,width=2.8in,bb=0 0 450
  420,clip]{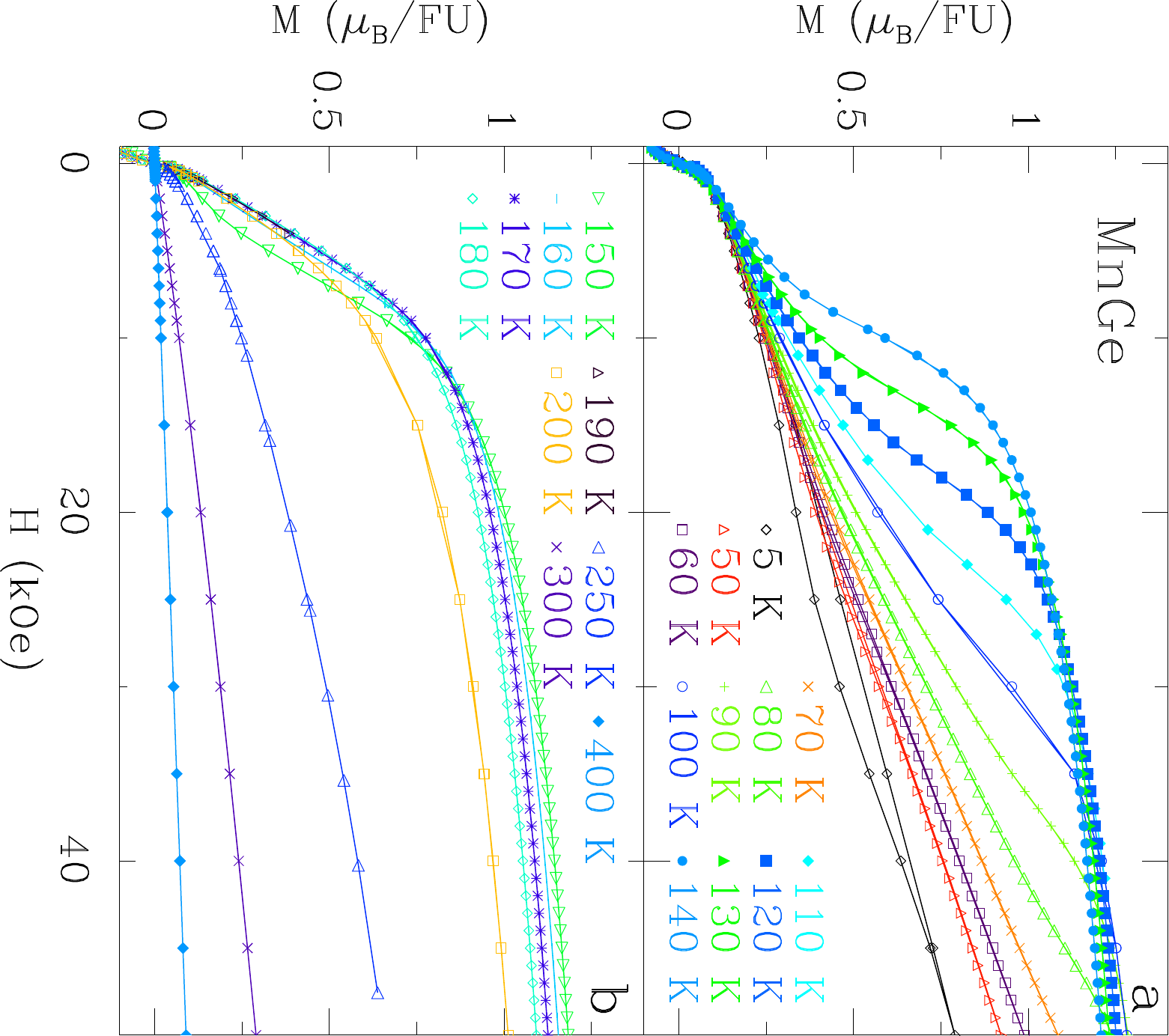}%
  \caption{\label{fig:magfig} Magnetization. Magnetic field, $H$,
dependence of the magnetization, $M$, for MnGe at temperatures between
5 and 140 K (a) and between 150 and 400 K (b) as indicated in the
figure. Lines connect the data points for display purposes. The
hysteresis at low temperatures is apparent in the data at 5 K where
both the initial increasing $H$ magnetization (lower data) and the
subsequent decreasing $H$-sweep (upper data) are shown. }
\end{figure}

\begin{figure}[bht]
  \includegraphics[angle=90,width=2.8in,bb=0 0 250
  410,clip]{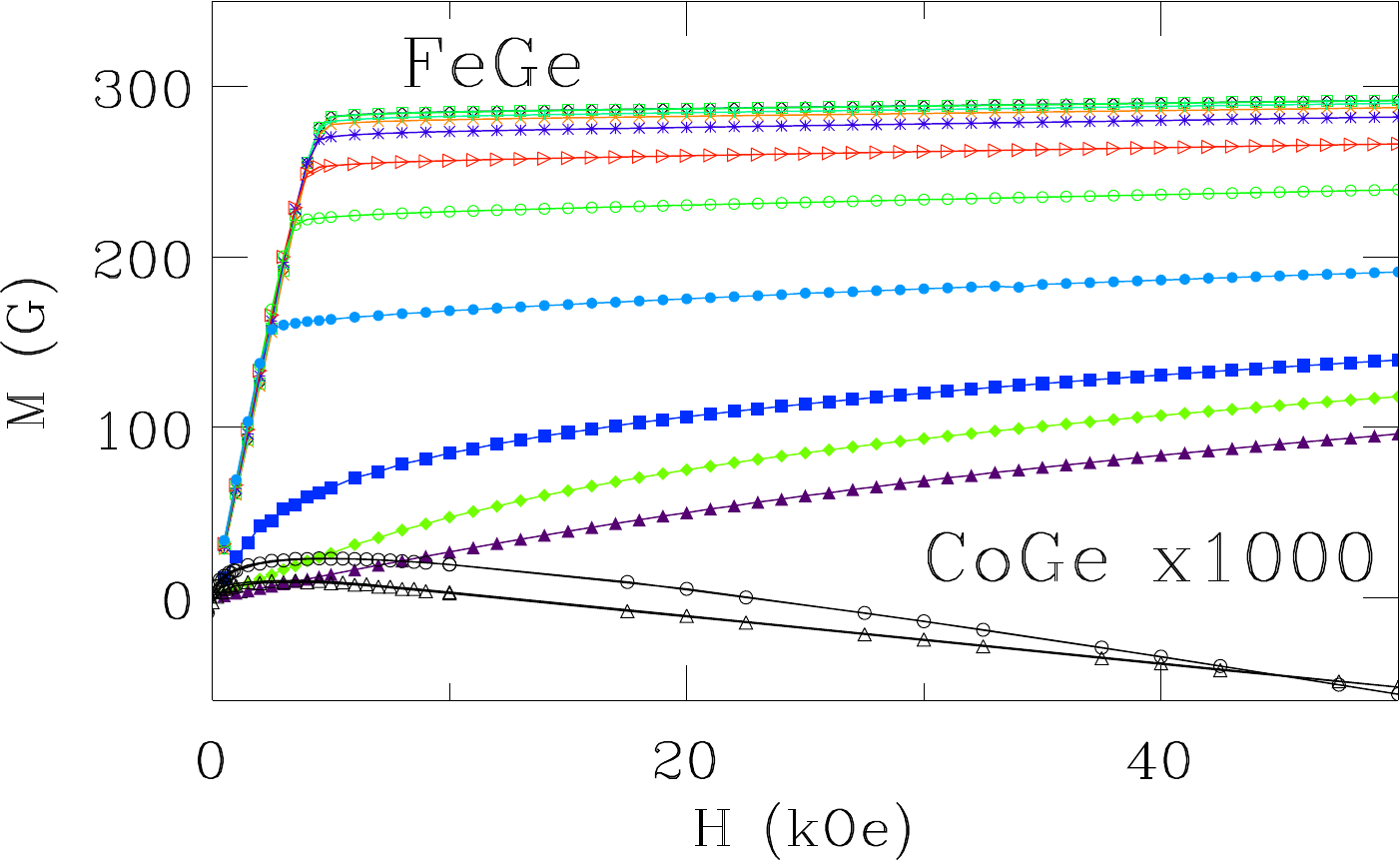}%
  \caption{\label{fig:magfigfege} Magnetization of FeGe and CoGe.
Magnetic field, $H$, dependence of the magnetization, $M$, for FeGe at
temperatures of 2 K (black diamonds), 5 K (blue triangles), 10 K
(green squares), 20 K (orange x's), 50 K (blue-green +'s), 100 K (blue
*'s), 150 (red rightward-pointing triangles), 200 K (green circles),
250 K (light-blue bullets), 280 K (dark-blue filled squares), 290 K
(light-green filled diamonds), and 300 K (violet filled triangles).
The magnetization of CoGe multiplied by a factor of 1000 for clarity
is also shown at temperatures of 5 K (black circles) and 300 K (black
triangles). Lines connect the data points for display purposes. }
\end{figure}

Equally as interesting is the evolution of the magnetization of MnGe
in $H$ and $T$ as displayed in Fig.~\ref{fig:magfig}. Here the linear
$M(H)$ at 300 K and above confirms the PM state inferred from
$\chi(T)$ while for $160 \le T < 250$ K a large low field contribution
indicates a FM ordering with little hysteresis. These $M(H)$ curves
are similar to that displayed by FeGe\cite{lundgren} as shown in
Fig.~\ref{fig:magfigfege} and MnSi\cite{wernick} below $T_c$ where a
steep increasing $M(H)$ at low $H$ is followed by a near saturation.
However, MnGe requires about twice the field to reach an apparent
saturation ($\sim 12$ kOe) at these $T$'s as compared to FeGe and
MnSi. The small low $H$ hysteresis in $M(H)$ and the similarity of
$M(H)$ with the other B20 compounds is consistent with, but not
sufficient to determine, a HM ordered state in MnGe. We note that
neutron diffraction measurements\cite{kanazawa,makarova} have detected
satellite peaks indicating a HM state in MnGe for $T< 170$ K.

\begin{figure}[bht]
  \includegraphics[angle=90,width=2.8in,bb=0 0 250
  410,clip]{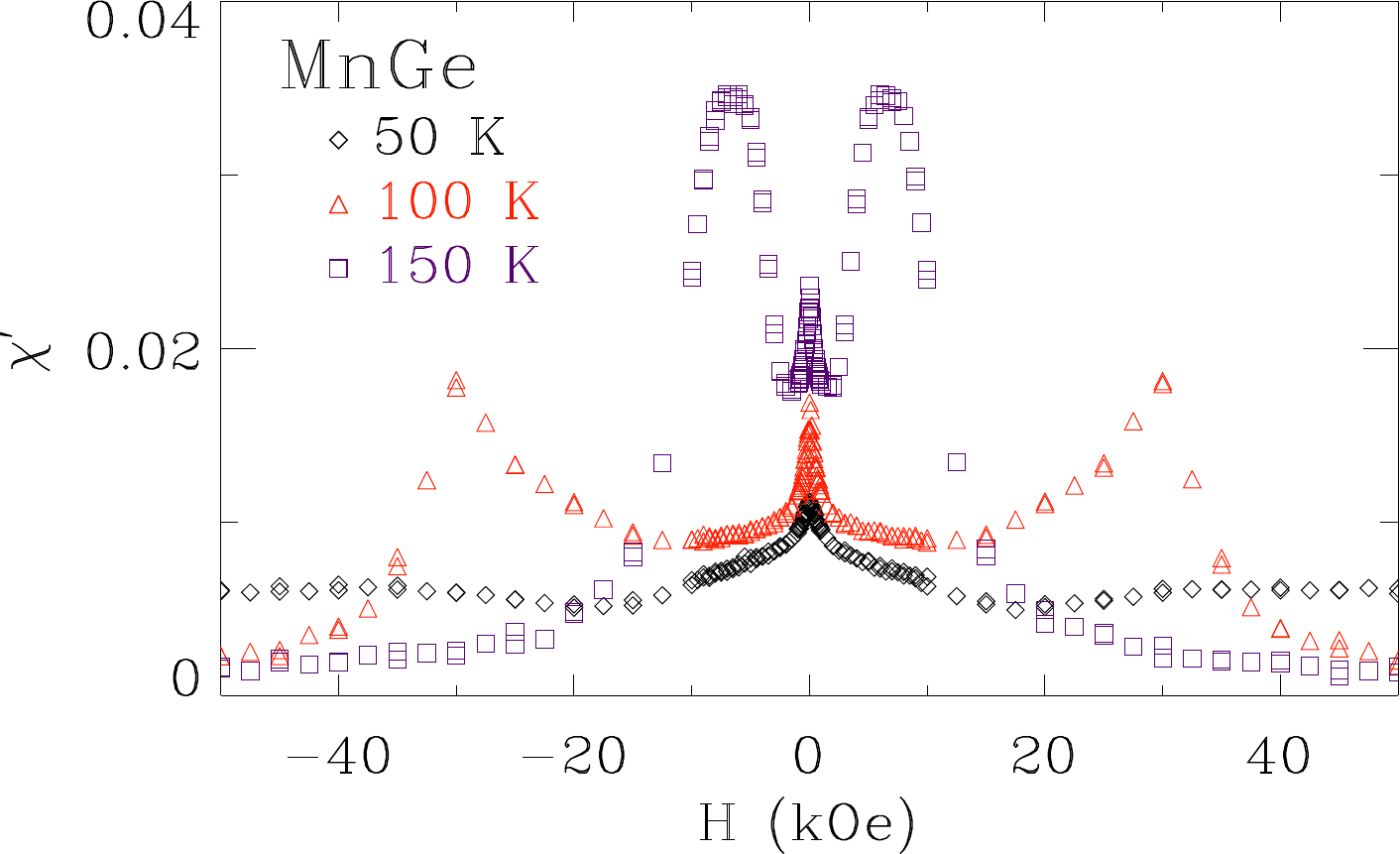}%
  \caption{\label{fig:acfsw} Field dependence of the ac
susceptibility. The real part of the ac susceptibility, $\chi'$ of
MnGe at three temperatures demonstrating the rich structure and sharp
transitions that occur with field, $H$, below the Curie
temperature. }
\end{figure}

In MnGe below $T=150$ K we observe an unusual suppression of $M(H)$
for $H > 1$ kOe with the saturation field moving to larger $H$ as $T$
is further decreased. These features are absent from the magnetization
data for FeGe in Fig.~\ref{fig:magfigfege}. For MnGe at $T< 70$ K we
are not able to access the saturation field as it has moved well above
our maximum field (50 kOe). We note that Kanazawa et al. have shown in
Ref.~\cite{kanazawa} that the saturation magnetization approaches 2
$\mu_B/$Mn at low temperatures and fields above 80 kOe. In addition to
the increased saturation field at lower $T$, we observe a strong
history dependence in the intermediate field range ($10 \le H \le 40$
kOe) as demonstrated at 5 K where we display both the initial up field
sweep after cooling in zero magnetic field and the subsequent downward
$H$- sweep.  This hysteresis is similar to what is measured in MnSi
and FeGe for fields between 1 and 6 kOe. Fig.~\ref{fig:acfsw}
displays the field dependence of $\chi'$ demonstrating the structure
and sharp transitions that occur with field. In particular, the very
sharp peak in the 100 K data near 30 kOe, as well as the broader peak in
the 150 K data below 10 kOe, correlate well with the sharp maxima found
in the $T$ dependence of Fig.~\ref{fig:chiacfig}a. These features occur
at fields a few kOe below the fields where $M$ begins to saturate. It
is interesting to note that detailed investigations of the ac
susceptibility of both MnSi and FeGe reveal a rich field dependence
that is restricted to temperatures near the Curie point. These have
been shown to be associated with the transitions into, and out of, the
$A$-phase that has been associated with a Skyrmion lattice state. In our
MnGe data we observe a somewhat different structure in $\chi'(H)$ over
a much wider $T$ range and at larger $H$.

The $M(H)$ for CoGe in Fig.~\ref{fig:magfigfege} shows a small
diamagnetic high field contribution, which may indicate that the
intrinsic behavior of CoGe is diamagnetic and that the low field
paramagnetic contribution may be extrinsic. Similar to the $T$
dependence of $\chi$ in Fig.~\ref{fig:chifig}, we observe only minor
$T$ dependence to the magnetization of CoGe. We do not observe any
features similar to that reported by Ref.~\cite{takizawa} in our CoGe
samples and find no indication of an antiferromagnetic transition near
120 K.

\begin{figure}[bht]
  \includegraphics[angle=90,width=2.8in,bb=0 0 250
  420,clip]{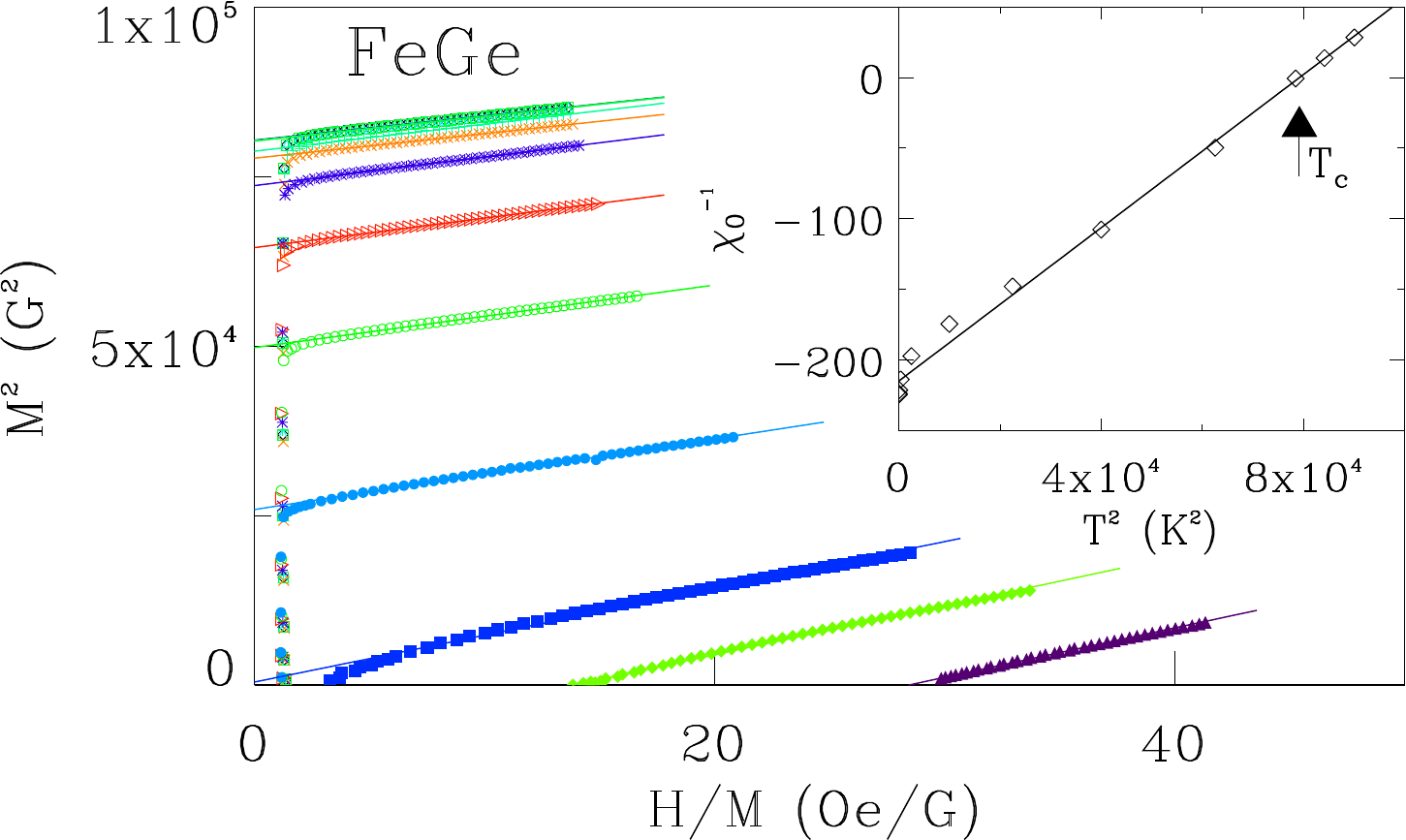}%
  \caption{\label{fig:arrottfege} Arrott plot for FeGe. The square of the
magnetization, $M^2$, plotted vs.\ the ratio of the magnetic field,
$H$, to $M$ to compare with a standard mean field form of $M$ commonly
known as an Arrott plot\protect{\cite{arrott}}. Symbols are the same
as in Fig.~\protect{\ref{fig:magfigfege}}. Solid lines are linear fits to
the high field behavior. Inset: Plot of the $H/M$
intercepts, $\chi_0^{-1}$, of linear fits in the main frame to
determine the Curie temperature, $T_c$, as indicated in the figure.}
\end{figure}

\begin{figure}[bht]
  \includegraphics[angle=90,width=2.8in,bb=0 0 450
  420,clip]{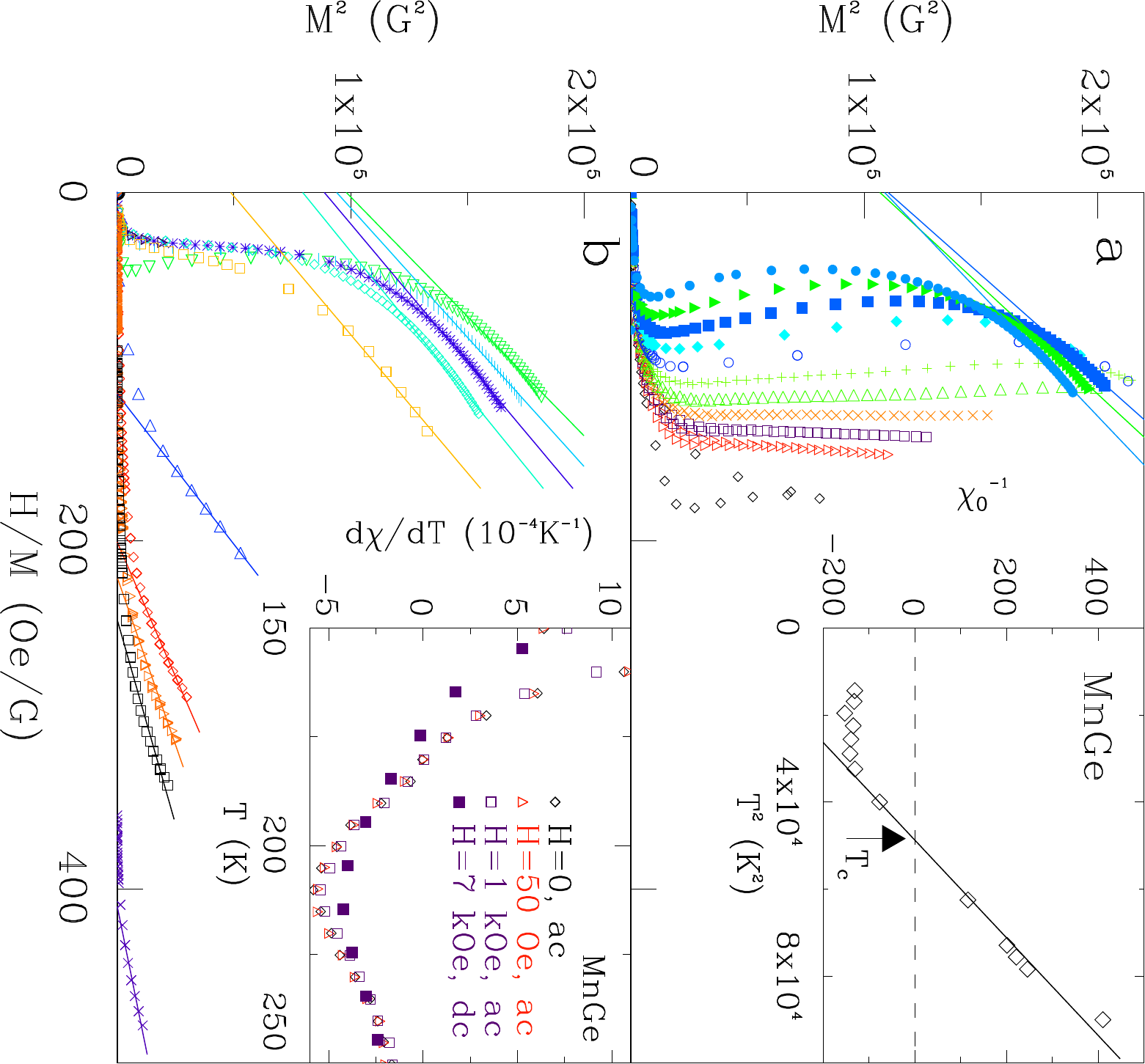}%
  \caption{\label{fig:arrott} Arrott plots for MnGe.  a) and b) the
square of the magnetization, $M^2$, plotted vs.\ the ratio of the
magnetic field, $H$, to $M$\protect{\cite{arrott}}. Symbols are the
same as in Fig.~\protect{\ref{fig:magfig}} with the addition of data
at 270 K (red diamonds), 275 K (orange triangles), and 280 K (black
squares) in frame b. Solid lines are linear fits to the high field
behavior. Inset to frame (a): Plot of the $H/M$ intercepts,
$\chi_0^{-1}$, of linear fits in the main frame to determine the Curie
temperature, $T_c$, as indicated in the figure.  Inset to frame (b):
derivative of the magnetic susceptibility, $\chi$, and $\chi'$ shown
in Figs.~\protect{\ref{fig:chifig}} and \protect{\ref{fig:chiacfig}},
with respect to temperature, $T$, vs.\ $T$. The minimum in $d\chi/dT$
corresponds closely to $T_c$ determined from the Arrott analysis in
the inset to (a).  }
\end{figure}

To begin to establish a magnetic phase diagram for MnGe, and to
compare directly with FeGe we have performed a standard mean field
analysis to our $M(H,T)$ data commonly known as an Arrott
plot\cite{arrott} as demonstrated in Fig.~\ref{fig:arrottfege} and
\ref{fig:arrott}. It is apparent from the linearity of $M^2$ as a
function of $H/M$ at high fields that FeGe is well described in terms
of a simple mean field treatment. In addition, the Curie temperature,
$T_c$, determined by finding the temperature where our high field
linear fits intercept the origin, agrees well with $T_c$ indicated by
the peak in $\chi(T)$ and with $\theta_W$.  In contrast the Arrott
plots for MnGe (Fig.~\ref{fig:arrott}) are not as simple to
interpret. We observe a change from a positive intercept of the $H/M$
axis, $\chi_0^{-1}$, for $T\ge 250$ K to a negative intercept at $T\le
200$ K indicating that $T_c$ lies between these two temperatures. We
locate the transition further by plotting $\chi_0^{-1}$ vs.\ $T^2$
displayed in the inset to Fig.~\ref{fig:arrott}a as suggested by the
Stoner-Edwards-Wohlfarth model for weak itinerant
ferromagnets\cite{stoner}. The plot indicates that $\chi_0^{-1}$ goes
to zero at $220\pm10$ K.  This is in contrast to both the peak in
$\chi(T)$ and $\chi'(T)$, and $\theta_W$ for MnGe which are consistent
with a magnetic transition $T$ near 275 K. However, as demonstrated in
the inset to Fig.~\ref{fig:arrott}b, both $d\chi/dT$ and $d\chi'/dT$
display a minimum at 210 K. Furthermore, $\chi''(T)$ shown in
Fig.\ref{fig:chiacfig}b displays a strong signal near 210 K at fields
above 20 kOe consistent with this critical temperature
determination. A more detailed modified Arrott analysis was carried
out to determine the variation of the critical temperature with the
choice of critical exponents. We found a variation of $T_c$ of no more
than 20 K for a wide range of exponents. Thus, we have observed an
inconsistency in the low field behavior indicating a magnetic
transition at 275 K and a standard Arrott analysis of the
magnetization above 20 kOe where our fits to the mean field form are
performed. We note that a previous investigation of MnGe has reported
a $T_c$ of 170 K\cite{kanazawa}, well below either of these transition
temperatures observed in our data.

The data presented here, including our $\chi'$ measurements of
Fig.~\ref{fig:chiacfig}a and Fig.~\ref{fig:acfsw}, and our $M(H)$ curves
below 150 K, suggest that the steep increase in $M(H)$ that we observe
at intermediate temperatures and fields may indeed be a thermodynamic
phase transition rather than a simple crossover between two simply
related magnetic states of MnGe. The wide field range required for
saturation of $M$ which increases upon cooling is clearly distinct
from that seen in FeGe and MnSi\cite{wernick,lundgren}. We note that
in this $T$ range Ref.~\cite{kanazawa,kanazawa2} report a significant
decrease in the HM period consistent with a larger saturation
field. The steep increase in $M(H)$ and the corresponding sharp peak
in $\chi'$ that we observe is interpreted as a phase transition just
below the field necessary to induce a ferromagnetically aligned state.
As such, this phase transition may be associated with a change from
the $A$-phase, where a Skyrmion lattice is expected, and a conical or a
field induced FM phase as occurs only in a small $T$-region below
$T_c$ in MnSi, FeGe, and Fe$_{1-x}$Co$_x$Si.

Our MnGe$_{1.2}$ samples displayed a $\chi$ and $M$ that reproduced
all of the essential features outlined for MnGe above. However, the
low field ($H\le 1$ kOe) $M$ was enhanced beyond that of MnGe by
almost a factor of 2 for $T < 275$ K, while the 50 kOe results were
within error of each other. In addition, like our CoGe sample, our
CoGe$_{1.1}$ sample displayed a small, $T$-independent, $\chi$ of
$6\times 10^{-6}$ from 400 down to 50 K with a Curie tail apparent at
low-$T$. The Curie-like behavior observed below 20 K is consistent
with a spin 1/2 impurity concentration of 1\%.

\subsection{B. Specific Heat}
\begin{figure}[bth]
  \includegraphics[angle=90,width=3.2in,bb=0 0 450
  360,clip]{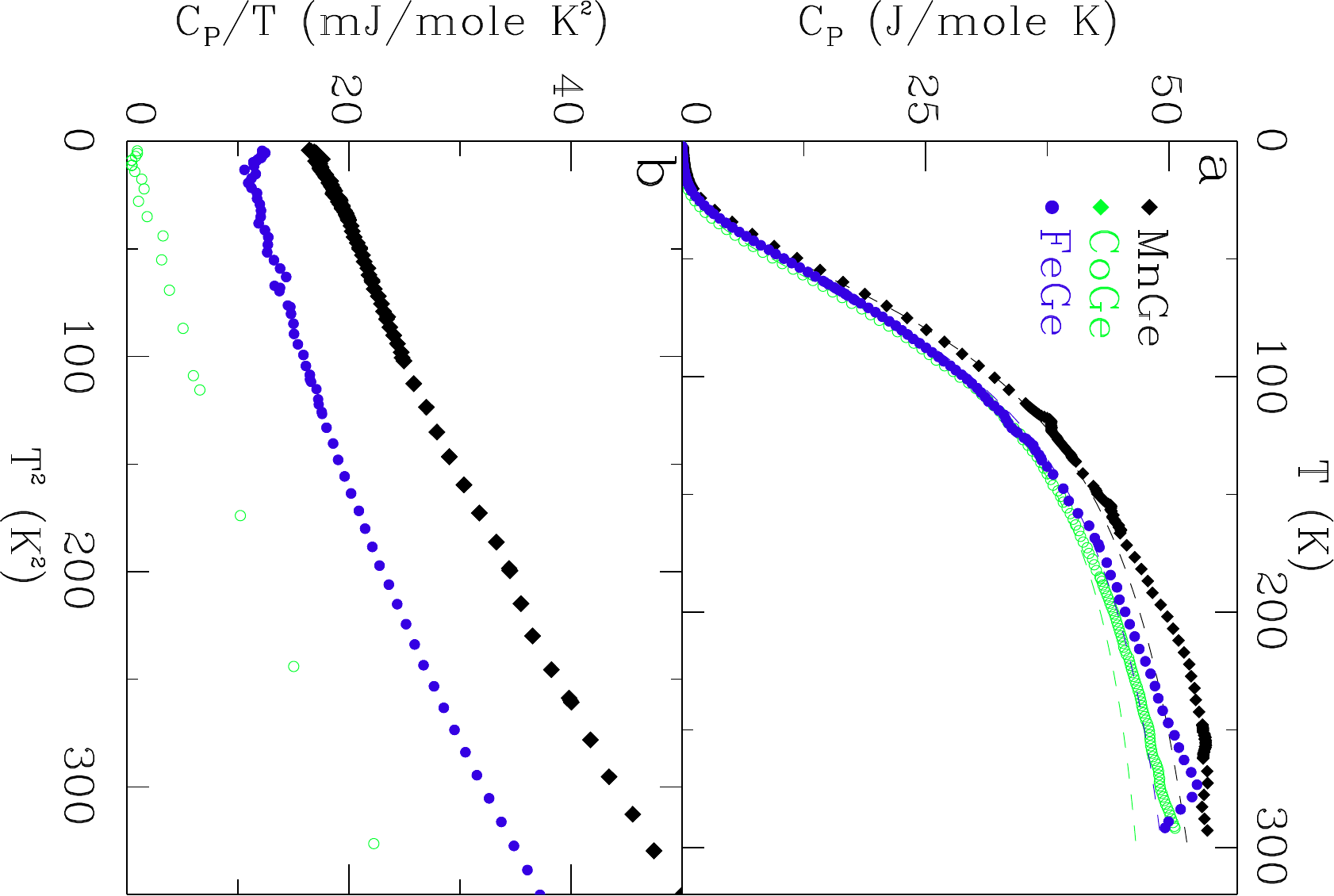}%
  \caption{\label{fig:sphfig1} Specific Heat. (a) Specific heat,
$C_P$, vs.\ temperature, $T$, of MnGe, FeGe, and CoGe at zero field.
Dashed lines are fits of a model for the phonon contribution to $C(T)$
that includes Debye and Einstein terms. (b) $C_P/T$ vs.\ $T^2$ at zero
field and low temperatures.}
\end{figure}

The specific heat, $C_P$, at $H=0$ of MnGe, CoGe, and FeGe is plotted
in Fig.~\ref{fig:sphfig1}a for $2.0 \le T \le 300$ K. Small peaks at
119, 160, and 165 K as well as significant structure above 200 K are
apparent in the MnGe data. In order to parametrize these data, we have
fit the Debye model with Einstein terms added to represent the optical
phonon contribution to our data\cite{ashcroft}. This model with Debye
temperatures of 272, 269, and 281 K and characteristic $T$ of 324,
282, and 319 K for the optical branches for FeGe, MnGe, and CoGe
respectively, represents the data fairly well. A clear deviation of
the model from the data can be seen at high temperatures for all 3
compounds as well as the peak structures noted above for MnGe and the
peak at the Curie temperature of FeGe. Frame (b) of the figure shows
$C_P/T$ plotted as a function of $T^2$ in the usual manner for
displaying the electronic and spin wave contributions to $C_P$.  Here,
it is apparent that the linear-in-$T$ terms are quite different for
the three compounds with coefficients, $\gamma$, of 16,
9\cite{yeo,marklund}, and $\sim$0 mJ / mole K$^2$ for MnGe, FeGe, and
CoGe respectively. $\gamma$ of MnGe and FeGe are comparable to that
found in MnSi (32 mJ / mole K$^2$)\cite{bauer}.

\begin{figure}[bth]
  \includegraphics[angle=90,width=3.2in,bb=0 0 450
  360,clip]{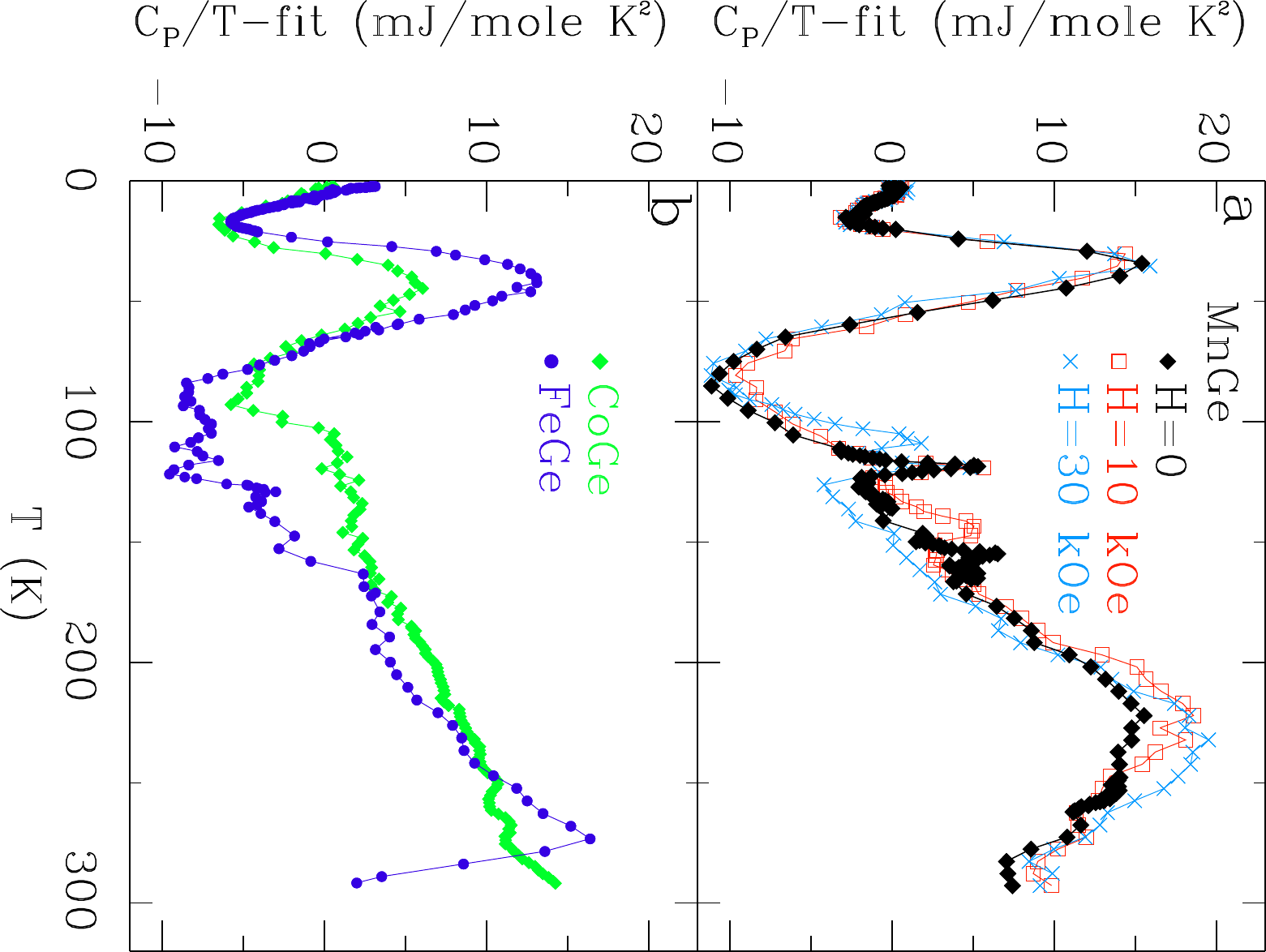}%
  \caption{\label{fig:sphfig2} Specific Heat after subtraction of a
model for the phonon contributions. (a) $C_P / T$ of MnGe at fields
identified in the figure after subtraction of the model represented by
the dashed line in Fig.~\protect{\ref{fig:sphfig1}}(a) vs.\
$T$. Several phase transitions are apparent. (b) $C_P/T$ vs.\ $T^2$
for FeGe and CoGe at zero field after subtraction of the model
represented by the dashed lines in
Fig.~\protect{\ref{fig:sphfig1}}(a). Lines connect the data points for
display purposes. }
\end{figure}

In Fig.~\ref{fig:sphfig2} (a) and (b) we plot $C_P / T$ after
subtracting the model represented by the dashed lines in
Fig.~\ref{fig:sphfig1}(a) in order to highlight the more subtle
features of $C_P/T$. $C_P/T$ for MnGe is plotted for fields of 0, 10,
and 30 kOe in Fig.~\ref{fig:sphfig2}a to display its field
dependence. We observe several features that these data have in common
for the three materials, including a broad maximum near 50 K that we
attribute to phonon contributions that are not reproduced by our
simple model. In addition, there are several features that are
apparent in the MnGe and FeGe data that do not appear in our CoGe
data. For FeGe we observe a sharp peak in $C_P(T)$ at the Curie
temperature which similar to that measured in Ref.~\cite{wilhelm2}.
The specific heat of MnGe appears to be far more interesting
displaying several distinct peaks and a broad region above 200 K with
a magnitude that exceeds our simple model of the phonon contribution
by $\sim 10$ mJ/mol K$^2$.  However, no distinct sign of a phase
transition near 275 K is visible in our data. Perhaps more interesting
are the 3 sharp peaks in the $H=0$ data of MnGe, one at 119 K and a
pair at 160 and 165 K, close to the magnetic transition temperature
identified in Ref.~\cite{kanazawa}. Application of a magnetic field of
up to 30 kOe has no effect on the temperature or shape of the peak we
observe at 119 K. In contrast, we observe a steep decrease in the
peak-$T$ of the 160 K feature with $H$ so that a single peak is
observed at 145 K in 10 kOe and at 110 K at 30 kOe. Thus, by 30 kOe
the field dependent peak occurs at lower $T$ than the field
independent, $T=119$ K, feature.

The field dependent phase transition identified here corresponds
closely in $T$ and $H$ with the peaks we observed in $\chi'$ in
Fig.~\ref{fig:chiacfig} and $dM/dH$ determined from the data in
Fig.~\ref{fig:magfig} and which we associate with a phase transition
that occurs at several kOe below the field needed to induce a FM
state. In contrast, we see no indication in any of the magnetic
measurements for a phase transition at 119 K, which, when considered
along with the observation that this transition is not field
dependent, suggests that this transition may not be related to the
magnetic state of the system.  It is somewhat difficult to reconcile a
phase transition in either the structure or electronic properties of
this itinerant magnet that has no effect on the magnetic
properties. As we noted above, we have measured the single crystal
X-ray diffraction of a small crystal separated from the melt at $T$s
down to 95 K in order to search for structural changes that may be the
cause of the 119 K phase transition. We found no differences in the
X-ray data outside of a 1\% thermal contraction of the lattice when
compared to our room $T$ results. It remains possible that there is a
more subtle structural transition that we are not sensitive to in our
single crystal X-ray diffraction experiments such as the tetragonal
distortion suggested in Ref.~\cite{makarova}. Thus far, we have not
identified the cause of the phase transition apparent in $C_P$ at 119
K in our MnGe sample.

\subsection{C. Resistivity}
\begin{figure}[htb]
  \includegraphics[angle=90,width=2.8in,bb=0 0 250
  420,clip]{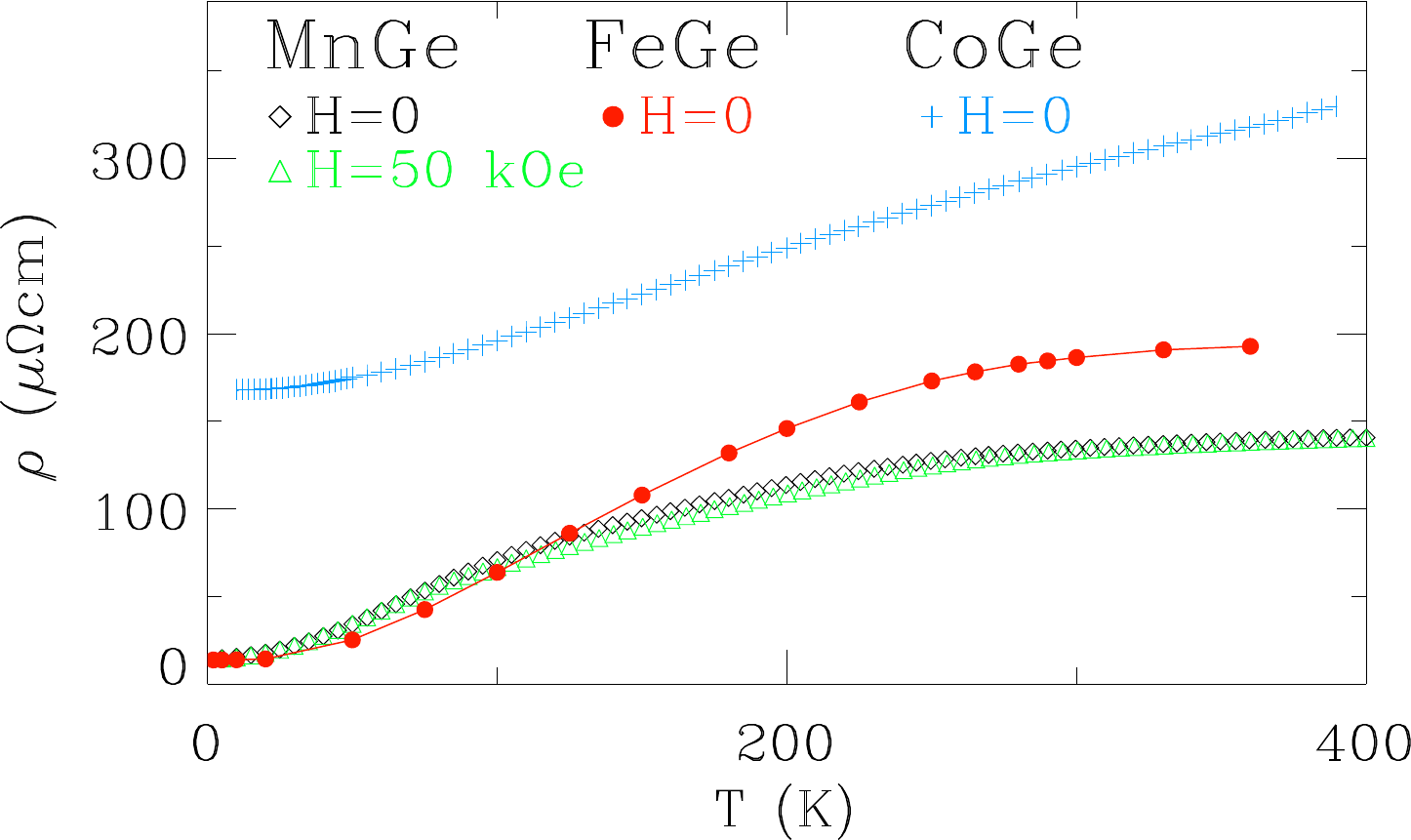}%
  \caption{\label{fig:resfig1} Resistivity.  Resistivity, $\rho$ vs.\
temperature $T$, at fields, $H$, identified in the figure for FeGe,
MnGe, and CoGe. }
\end{figure}

The resistivities of FeGe, MnGe, and CoGe displayed in
Fig.~\ref{fig:resfig1} are that of metals with residual resistivity
ratios, $RRR=\rho(300$ K$) / \rho(4$ K$)$, of 14, 9.3, and 1.8 for
FeGe, MnGe, and CoGe respectively. The corresponding residual
resistivities $\rho(4$ K$)$ are 13.5, 15.5, and 160 $\mu
\Omega$cm. There are no obvious features in the $T$ dependence of
$\rho$ that correspond to either $T_c$'s of FeGe and MnGe nor the
phase transition identified in $C(T)$ at lower $T$ for MnGe. Below 10
K a $T^2$ dependence of $\rho(T)$ becomes apparent in FeGe and MnGe
with a coefficients, $A = 0.8$ and $9.2 $ n$\Omega$cm/K$^2$.  These
values can be compared to that measured in MnSi which is about 3 times
larger than our value for MnGe\cite{capan,kadowaki2}. It is
interesting to note that for these compounds, FeGe, MnGe and MnSi, the
ratio $A/\gamma^2$ is between $1\times10^{-5}$ and $3.0\times10^{-5}
\mu\Omega$cm (mole K/mJ)$^2$ or about 1 to 3 times the value reported
by Kadowaki and Woods for heavy Fermion
metals\cite{bauer,kadowaki1}. In MnGe the application of a 50 kOe field
reduces $\rho$ by a few percent over most of the $T$ range covered
suggesting that the field reduces the magnetic fluctuation scattering
of the carriers.

\begin{figure}[htb]
  \includegraphics[angle=90,width=2.8in,bb=0 0 450
  420,clip]{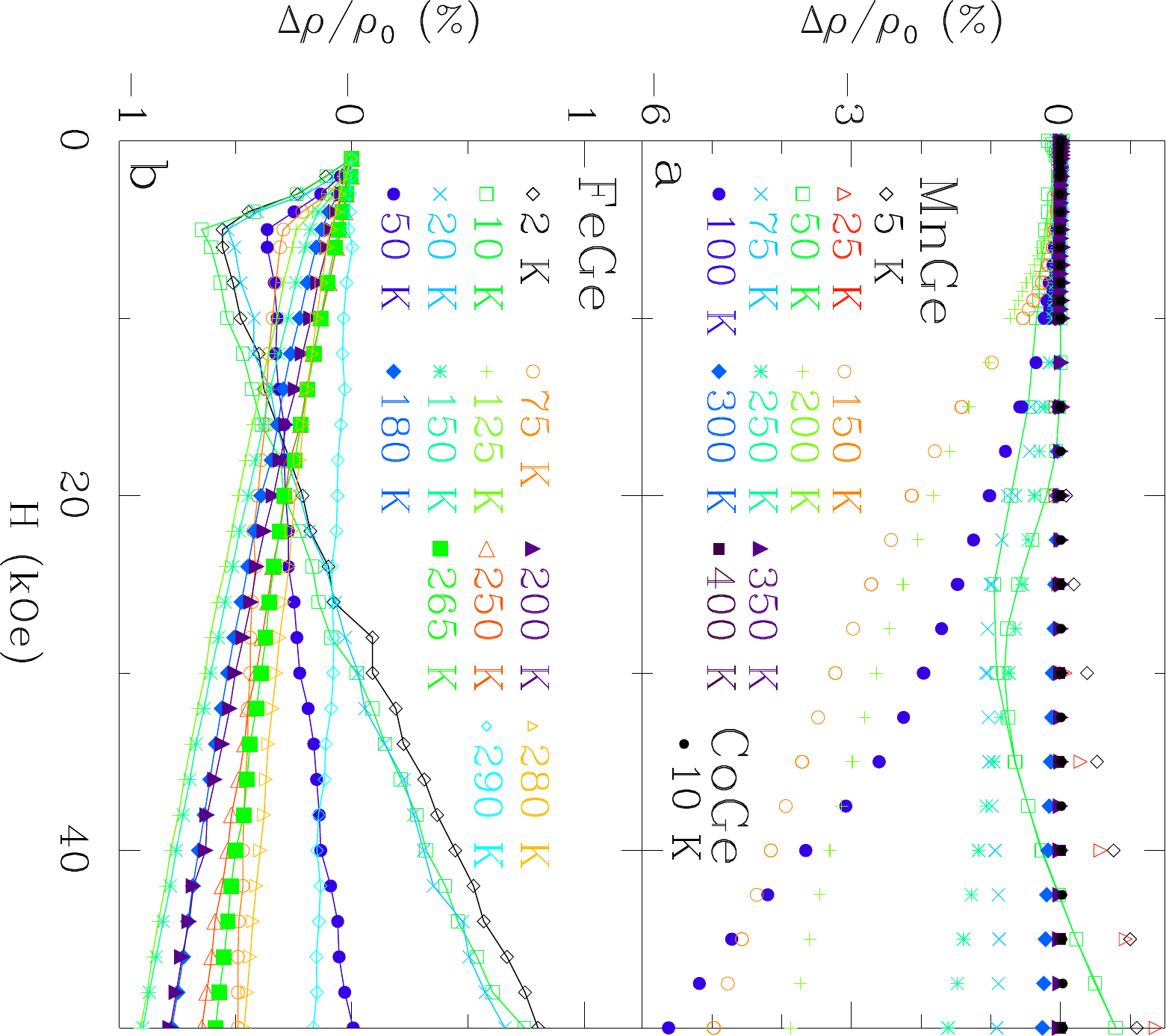}%
  \caption{\label{fig:mrfig} Magnetoresistance. (a) Magnetoresistance,
$\Delta\rho/\rho_0=(\rho(H)-\rho_0)/\rho_0$, where $\rho_0$ is the zero
field resistivity, of MnGe and CoGe at temperatures indicated in the
figure. The hysteresis at low temperatures is demonstrated by the 50 K
data where both the increasing and decreasing field data are
shown. (b) $\Delta\rho / \rho_0$ for FeGe at temperatures indicated in
the figure. Lines connect the data points for display purposes. }
\end{figure}

 A more detailed view of the magnetoresistance, MR, of all three
compounds is shown in Fig.~\ref{fig:mrfig}a and b where the field
dependence is displayed. For both MnGe and FeGe a negative $\Delta
\rho / \rho_0$, where $\Delta \rho = \rho(H) - \rho_0$ and $\rho_0$ is
the zero field resistivity, reaches its largest absolute value near
100 K.  At lower $T$ a positive MR contribution is observed that grows
with reduced $T$.  In FeGe a distinct change in the MR occurs at the
transition to the field induced FM state near 6 kOe. A similar, yet
less obvious change is visible in the MR of MnGe along with a
hysteretic behavior associated with the transition to the FM phase and
which is maximized near 50 K.  We demonstrate the hysteric MR in the
figure by displaying both the increasing and decreasing field data at
50 K. This hysteresis occurs in the same field range as the hysteresis
identified in $M(H)$ in Fig.~\ref{fig:magfig}.  These features of the
MR of MnGe and FeGe are similar to that observed in MnSi as all three
display a negative MR with sharp changes observed near the crossover
to field induced ferromagnetism. However, in MnSi the negative MR near
$T_c=30$ K can be as large as -40\% for clean samples (see
Ref.~\cite{leeprb} for example). The obvious hysteresis demonstrated
in the figure for MnGe appears to be unique to this compound. In all
three of these materials the positive contribution to the MR dominates
at $T<5$ K. The MR of CoGe is very small and positive being less than
0.1\% at 10 K and 50 kOe (Fig.~\ref{fig:mrfig}).

The carrier densities of all three compounds have been estimated from
the high field Hall effect. A detailed presentation of the Hall data
including the anomalous Hall effect\cite{kanazawa} will be presented
elsewhere. Here we report the carrier densities implied by the
simplest interpretation of the Hall constant $R_0 = 1/nec$ as
presented in frame d of Fig.~\ref{fig:tccomp}. While MnGe and FeGe
have metallic carrier densities consistent with about 1/2 to 2/3
carrier per formula unit, CoGe has a much smaller carrier density,
rising with $T$ from 0.6 to $1.3\times 10^{21}$ cm$^{-3}$. Thus,
while MnGe and CoGe appear to be very similar to their silicide
counterparts, MnSi and CoSi, in their magnetic and electronic states,
FeGe is well known to be very different from the non-magnetic
insulator FeSi\cite{yeo}.

\section{IV. Electronic Structure Calculations}

In order to better determine where electronic interactions are
significant in the transition metal ({\it TM}) germanides with the B20
crystal structure and to compare their relative importance, we
calculate the electronic properties of MnGe, FeGe, and CoGe using a
standard density functional approach. The B20 structure is a cubic
structure with a rather low internal symmetry, as demonstrated in
Fig.~\ref{fig:struc}, so that there is neither 4-fold rotation
symmetry nor inversion symmetry in the lattice. This low symmetry is
thought to be responsible for the interesting magnetic properties of
materials having this structure as outlined in the
introduction. Although the effects of the low symmetry, as reflected
in the importance of spin-orbit coupling, for example, may be
restricted to energy scales too small to be captured in our models,
our calculations are useful in that they provide a starting point for
understanding what role spin-orbit splitting may play. Similarly, this
method is a clear approximation in terms of the electron interactions
in the strongly localized $d$-orbitals. However, we believe that it
provides a very useful first approximation to understanding the role
the interactions play. By comparing these results with experiment, we
can identify where the electronic interactions and spin-orbit
splitting of the bands are significant and where they play a minor
role.  We also investigate these materials in a variety of conditions,
varying the lattice constant as might be accomplished with alloying to
examine the possible ground states and to compare to the experimental
results presented here.

We use the standard WIEN2K all-electron DFT package\cite{wien2k} that
uses a LAPW basis including local orbitals.  We have chosen for our
study to use the GGA functional\cite{gga}, which provides a
significant improvement over LDA in terms of bond lengths and other
properties.  It is also considered to be a good choice for systems of
moderate interaction.  For the calculation of MnGe, we chose muffin
tin radii of 2.34 a.u.\ for Mn and 2.23 a.u.\ for Ge.  For FeGe the
{\it TM} and Ge radii were 2.28 a.u.\ and 2.17 a.u., and for CoGe they
were 2.31 a.u.\ and 2.23 a.u.  For MnGe we used u=0.135 and 0.842, for
FeGe we used u=0.130 and 0.839, and for CoGe we used u=0.140 and
0.841.  These values are close to the experimental positions, in our
calculations we found the residual forces at these positions to be
less than 0.03 Ryd/au.  In the calculations, the plane wave cutoff
$R*K_{max}$ was varied from 7.0 to 9.0 to ensure the basis set and
energies had converged.  For most of the calculations, we employed a
grid of 19x19x19 $k$-points for Brillouin zone (BZ) integrations (340
in the irreducible zone).  For FS calculations we used a denser
34x34x34 grid.  We used the modified tetrahedron method\cite{fsinteg}
to perform integrals over the Brillouin zone.  While the observed
ground state of MnGe and FeGe is helimagnetic, the pitch of the
helical order is much larger than the unit cell size, so we have done
our magnetic calculations assuming a uniform ferromagnetic state.

\begin{figure}[htb]
  \includegraphics[angle=0,width=2.8in,bb=15 140 345
  500,clip]{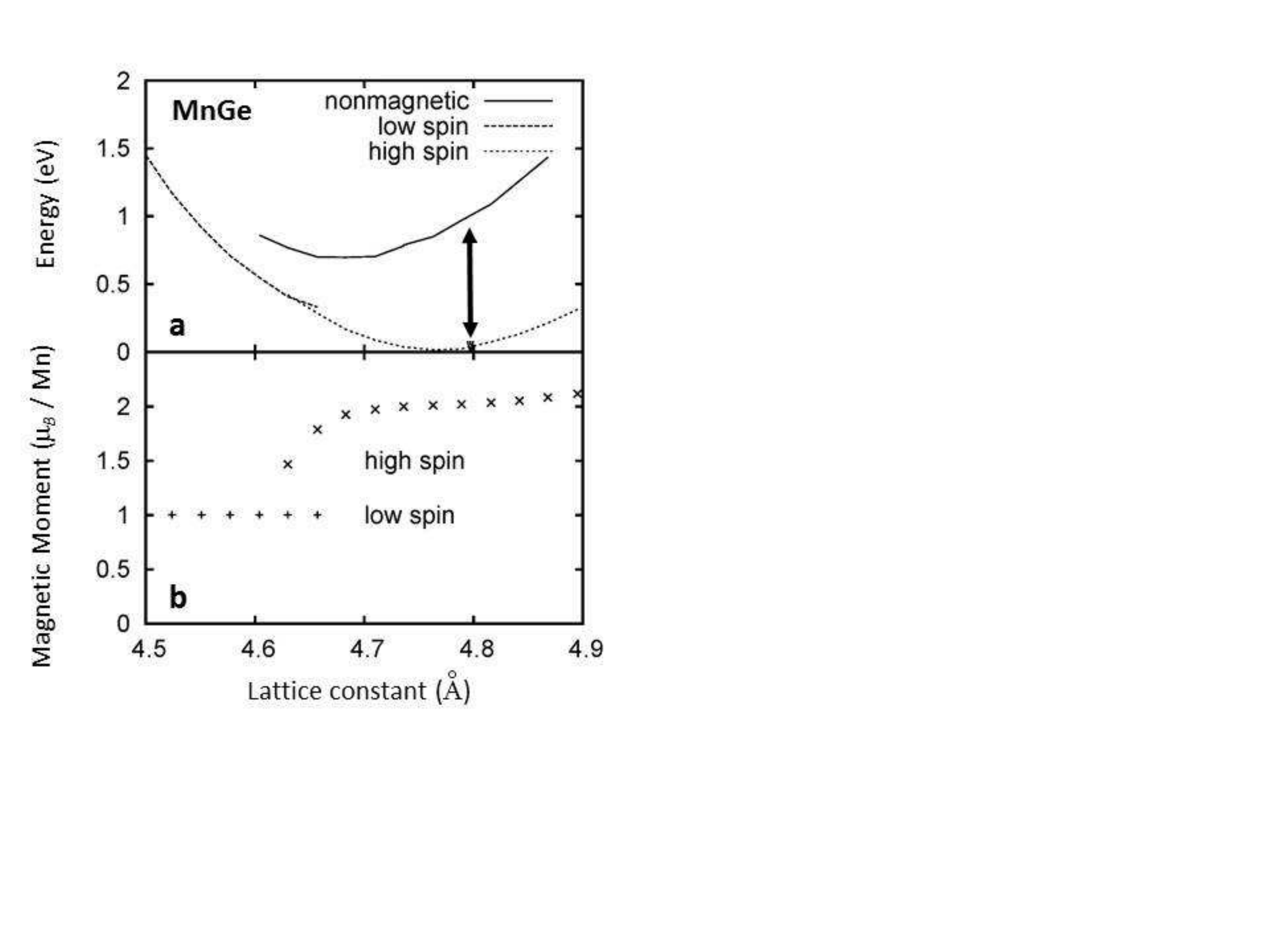}%
  \caption{\label{fig:bs1} Total Energy and Magnetic Moment of
MnGe. (a) Total energy per unit cell variation as a function of MnGe
lattice constant. Arrow indicates the experimental lattice
constant. (b) Magnetic moment per Mn atom calculated as a function of
lattice constant. }
\end{figure}

In Fig.~\ref{fig:bs1} we show the total energy and total magnetic
moment of MnGe in possible magnetic (2) and nonmagnetic ground states
for a range of lattice constants. We find that a magnetic ground state
is preferred at the experimental lattice constant, and remains that
way until the lattice constant is reduced to 4.46 A (not shown). There
appears to be two competing magnetic ground states in MnGe with
Fig.~\ref{fig:bs1}b showing that the predicted moment is close to 2
$\mu_B$/Mn at the experimental lattice constant and that the moment
abruptly collapses to 1 $\mu_B$/Mn at about 4.6 $\AA$ similar to what
was predicted in Ref.~ \cite{robler}. This drastic change in the
electronic structure suggests measurements of the magnetic moment and
magnetic ordering under applied pressure such as that recently carried
out by Deutsch {\it et al.}\cite{deutsch}. The calculated Bulk Modulus
of MnGe is 238 GPa so that a characteristic pressure for the
transition is predicted to be close to 28 GPa suggesting diamond anvil
techniques may be required to access this transition. However, in
Ref.~\cite{deutsch} that this transition was found to occur at a much
smaller pressure, near 6 GPa. Interestingly, the chiral order was
found to persist at pressure above the transition, but with a reduced
helical pitch.

\begin{figure}[htb]
  \includegraphics[angle=0,width=2.8in,bb=5 5 695
  500,clip]{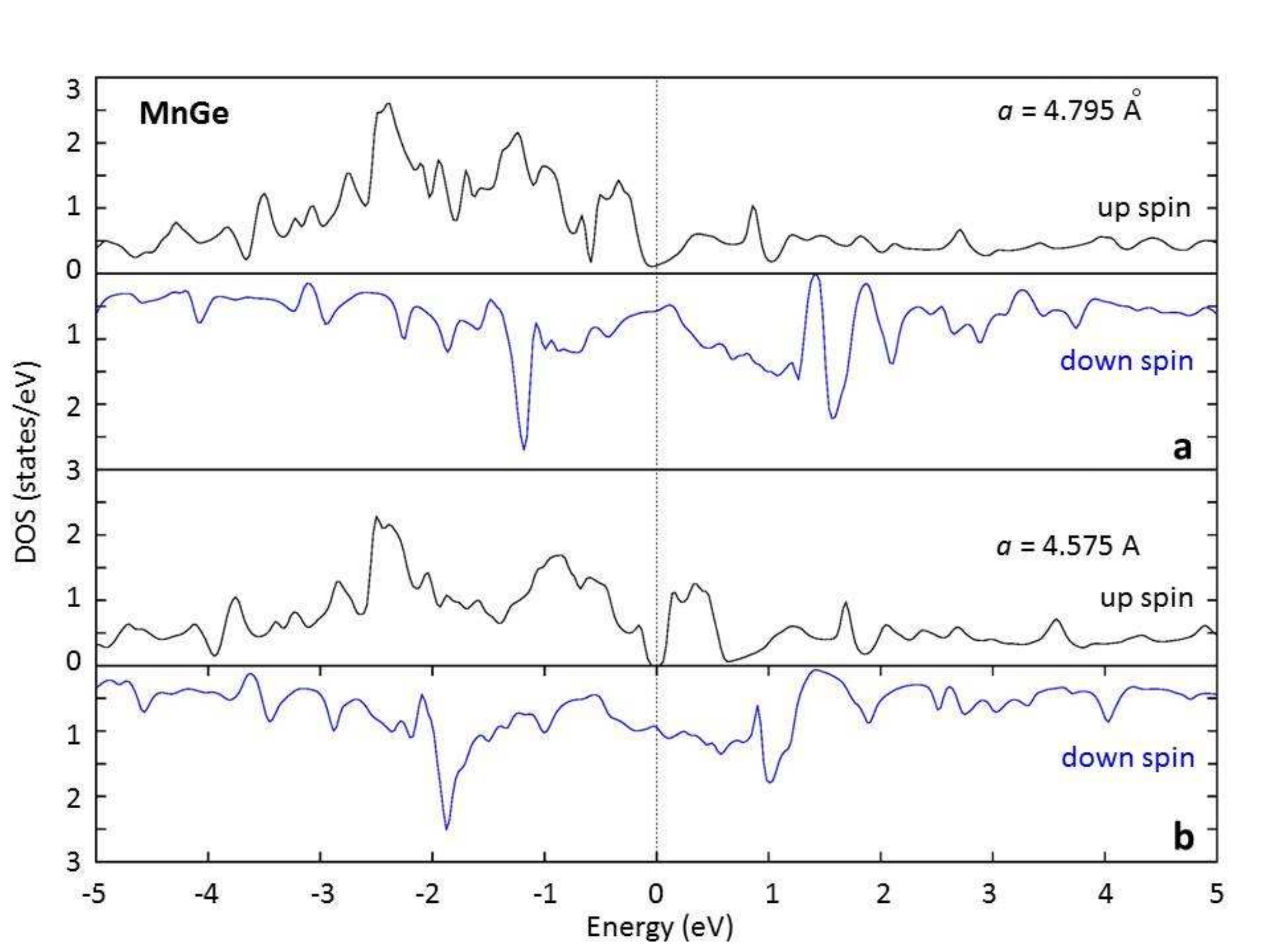}%
  \caption{\label{fig:bs3} DOS for MnGe for two different lattice
spacings. Density of states (DOS) for both the majority (spin up) and
minority (spin down) sub-bands calculated at the experimental lattice
constant (a) and at a smaller lattice constant (b). }
\end{figure}

The nature of the rapid collapse of the magnetic moment with reduction
of the lattice constant can be clearly seen in the DOS of MnGe
presented for two lattice spacings (4.795 $\AA$, high moment solution)
and (4.575 $\AA$, low moment solution), shown in Fig.~\ref{fig:bs3}.
The 2 $\mu_B$/Mn state has both spin-up and spin-down carriers present
at the Fermi energy, but the low moment state seen at small lattice
constant is a half metallic state with only minority spin carriers
present at the Fermi level\cite{robler}.  The predominant change is
that the peak in the spin-up DOS lying between -0.5 and 0 eV for the
high moment solution (Fig.~\ref{fig:bs3}a) has moved upward in energy
so that the gap visible near -0.5 eV in Fig. 3(a) moves to the Fermi
level in the low-spin solution shown in Fig.~\ref{fig:bs3}b. This
change corresponds to two complete bands shifting from just below the
Fermi energy to just above the Fermi energy. Hence we conclude that
under pressure, MnGe should become half-metallic and suffer a rapid
drop in total moment.  It is interesting to note that in
Fig.~\ref{fig:magfig} $M(H)$ tends to saturate at high $T$ at $M\sim
1.25 \mu_B$/Mn where as at low $T$ $M(H)$ tends to saturate near 2
$\mu_B$/Mn\cite{kanazawa}.  The mechanism for this variation with $T$
unknown, and our calculations do not explain why this change occurs
for such a small increase in $T$. We speculate, however, that the
variation in the saturated magnetization with $T$ may be related to
the competition between these two ground states that lie close in
total energy.

\begin{figure}[htb]
  \includegraphics[angle=0,width=2.8in,bb=60 0 455
  520,clip]{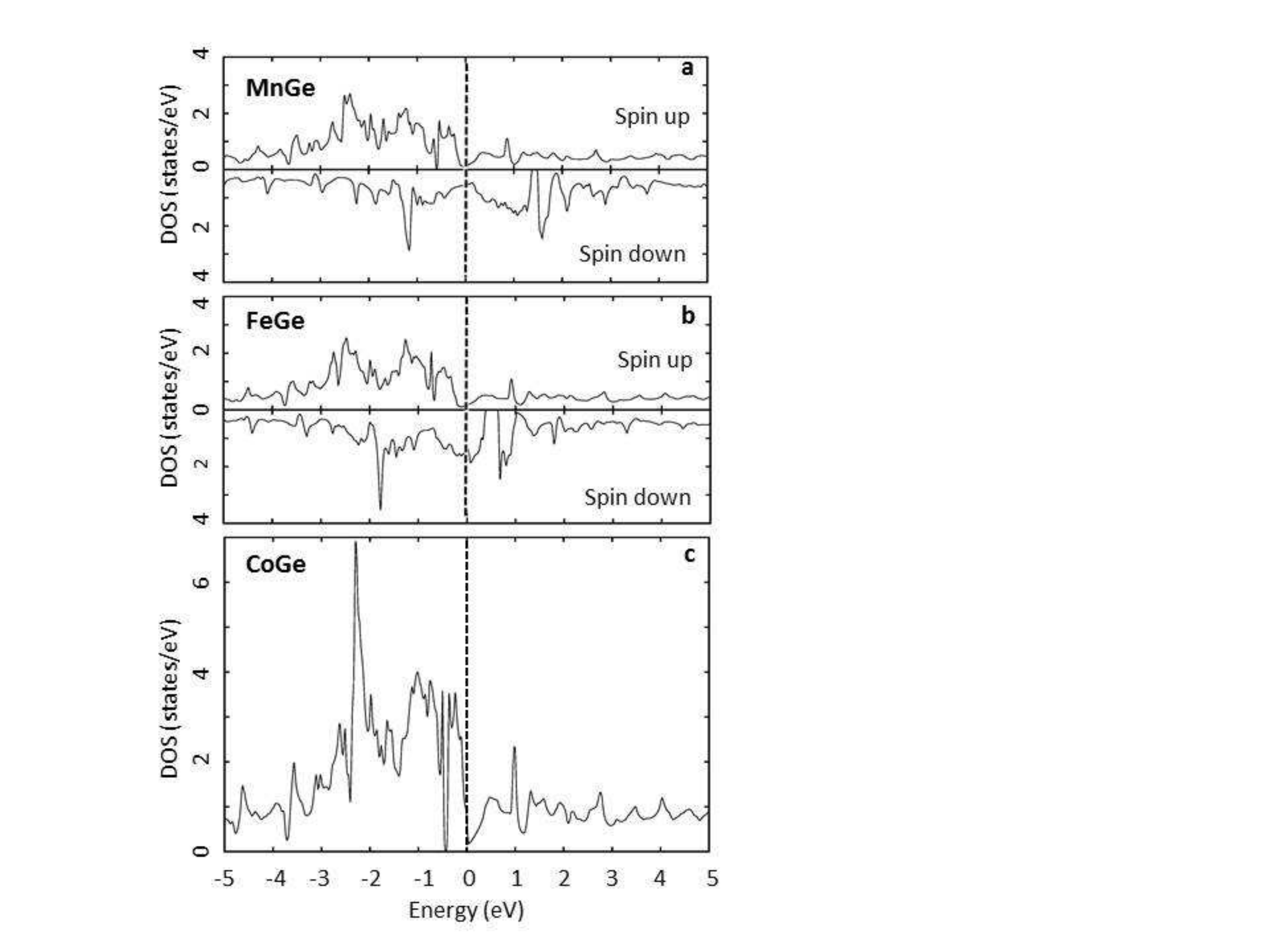}%
  \caption{\label{fig:bs2} DOS for MnGe, FeGe, and CoGe. Density of
states (DOS) for both the majority (spin up) and minority (spin down)
sub-bands of MnGe and FeGe, as well as the DOS of CoGe calculated at
the experimental lattice constants.  }
\end{figure}

In all of these B20 materials, including the silicides, the states
within 5 eV of the Fermi energy are predominately derived from the
{\it TM} d-orbitals, and it is believed they drive much of the
interesting properties of these compounds. To highlight the changes to
the DOS with filling of the $d$-orbitals across the series, and to
place our MnGe results in context, we show in Fig.~\ref{fig:bs2} the
total DOS for MnGe, FeGe and CoGe. What is most striking about this
figure is the similarity of the DOS in the majority spin bands of MnGe
and FeGe to the DOS of CoGe. In fact, the shape of the density of
states is similar across this series with the main difference being
the placement of the Fermi level. One remarkable feature made clear by
this plot is the appearance of a pair of gaps (or in some cases deep
minima) in the DOS within about 2 eV of the Fermi level. The gap that
appears at roughly 0.5 eV lower in energy is responsible for the well
studied insulating behavior of FeSi and where the large DOS seen at
the gap edges is thought to be important in creating many of its
interesting temperature and doping dependent features.  These
features, resulting from $d$-band splitting by the low symmetry of the
crystal field, are seen in almost all DOS of these {\it TM} silicides
and germanides\cite{yamada,jarlborg,kanazawa3,mazurenko,pan,jeong}.
We note that a magnetic ground state for FeGe is found with a magnetic
moment near 1 $\mu_B$ in agreement with prior calculations and
experiments\cite{lundgren,yamada,jarlborg,anisimov}. This is in
contrast to our calculations for CoGe in which the energy minimization
converges to nonmagnetic state for all lattice constants
probed\cite{kanazawa3}.  The defining feature of the DOS of CoGe is
the pseudo-gap residing close to the Fermi level in a similar fashion
to the majority band case of both MnGe and FeGe as well as to the case
of CoSi\cite{pan}. This placement of the Fermi energy reduces the DOS
at the Fermi level for CoGe to about 1 state per eV per unit cell. The
paramagnetic ground state of CoGe may very well be a result of a
smaller DOS at the Fermi level as compared to either the itinerant
magnets FeGe or MnGe. That the Fermi level lies in a deep valley in
the DOS also correlates well with the small apparent charge carrier
concentration, about 0.04 charge carriers per CoGe formula unit,
determined from the Hall voltage.

\begin{figure}[htb]
  \includegraphics[angle=0,width=3.4in,bb=50 40 550
  370,clip]{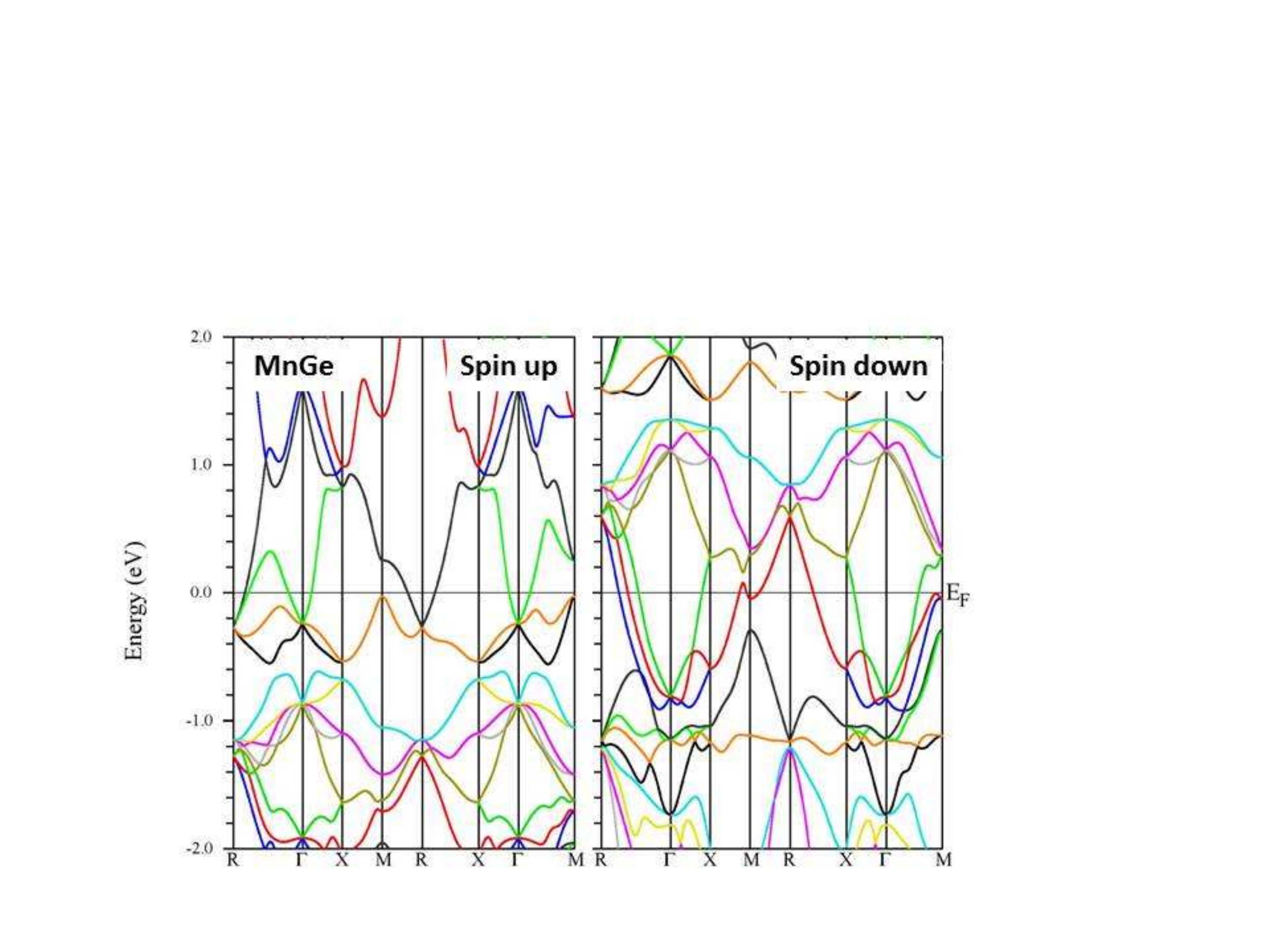}%
  \caption{\label{fig:bs4} Band structure of the majority (spin up)
and minority (spin down) sub-bands of MnGe. Energy bands calculated at
points of high symmetry at the equilibrium lattice constant. }
\end{figure}

\begin{figure}[htb]
  \includegraphics[angle=0,width=3.4in,bb=40 50 540
  355,clip]{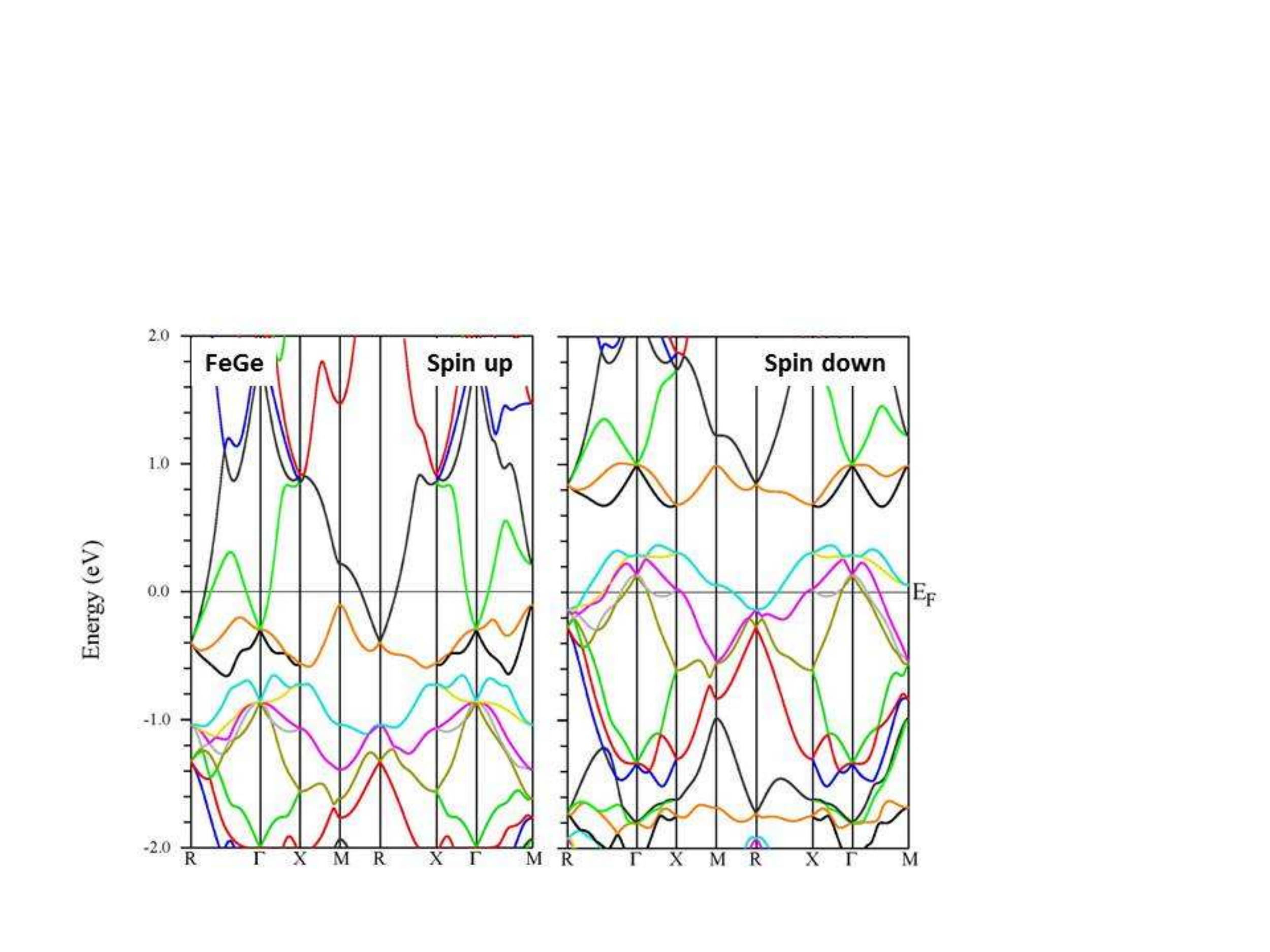}%
  \caption{\label{fig:bsfege} Band structure of majority (spin up) and
minority (spin down) spin sub-bands of FeGe. Energy bands calculated
at points of high symmetry at the experimental lattice constant. }
\end{figure}

\begin{figure}[htb]
  \includegraphics[angle=0,width=2.8in,bb=130 50 670
  520,clip]{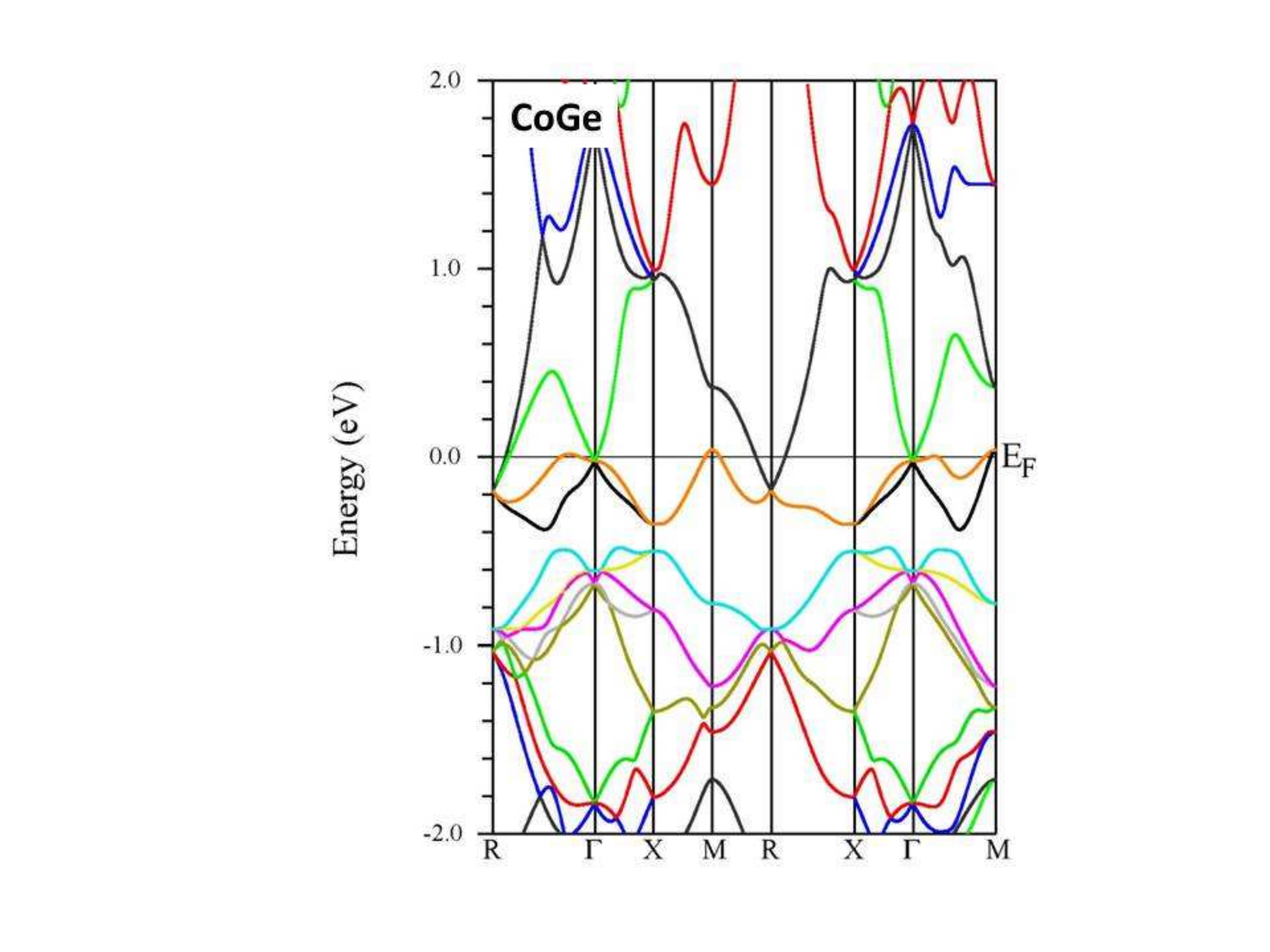}%
  \caption{\label{fig:bscoge} Band structure of CoGe. Energy bands
  calculated at points of high symmetry. }
\end{figure}

The similarity in the DOS of the three materials suggests that the
band structure can be well understood from a rigid band approximation
where only the electron count and the splitting of the Fermi energy in
the spin sub-bands accounts for the changes. This can be seen in
Figs.~\ref{fig:bs4}, \ref{fig:bscoge} and \ref{fig:bsfege}, where the
results of our band structure calculations are presented.  For MnGe
(Fig.~\ref{fig:bs4}) and FeGe (Fig.~\ref{fig:bsfege}) the band
structure is shown for both spin orientations and the identity of each
of the bands is indicated by its color so that they can be traced
across each of the 3 figures.  Again, we draw attention to the
similarity of the electronic structure of CoGe with the majority spin
bands in MnGe and FeGe. In each case the Fermi level lies near the top
of a set of two bands, identified by their black and gold colors, and
at the bottom of second pair of bands which are green and gray. In
addition, there appears to be a Dirac point in the electronic
structure of CoGe at the $\Gamma$-point, however, the consequences of
this unusual structure are likely masked by the larger population of
charge carriers associated with bands crossing the Fermi energy at
different $k$-points within the BZ\cite{kanazawa3}. A comparison to
the band structure of the silicides reveals many similarities with the
largest difference coming from the 30\% reduction in band width of the
germanides resulting from the $\sim 5$\% increase in lattice constants
(see Fig.~\ref{fig:tccomp}). As a result, the relative shift of the
spin sub-bands in MnGe and FeGe are larger than their silicide
relatives leading to the larger magnetic moments found in the
germanides.

\begin{figure}[htb]
  \includegraphics[angle=0,width=3.2in,bb=70 260 470
  470,clip]{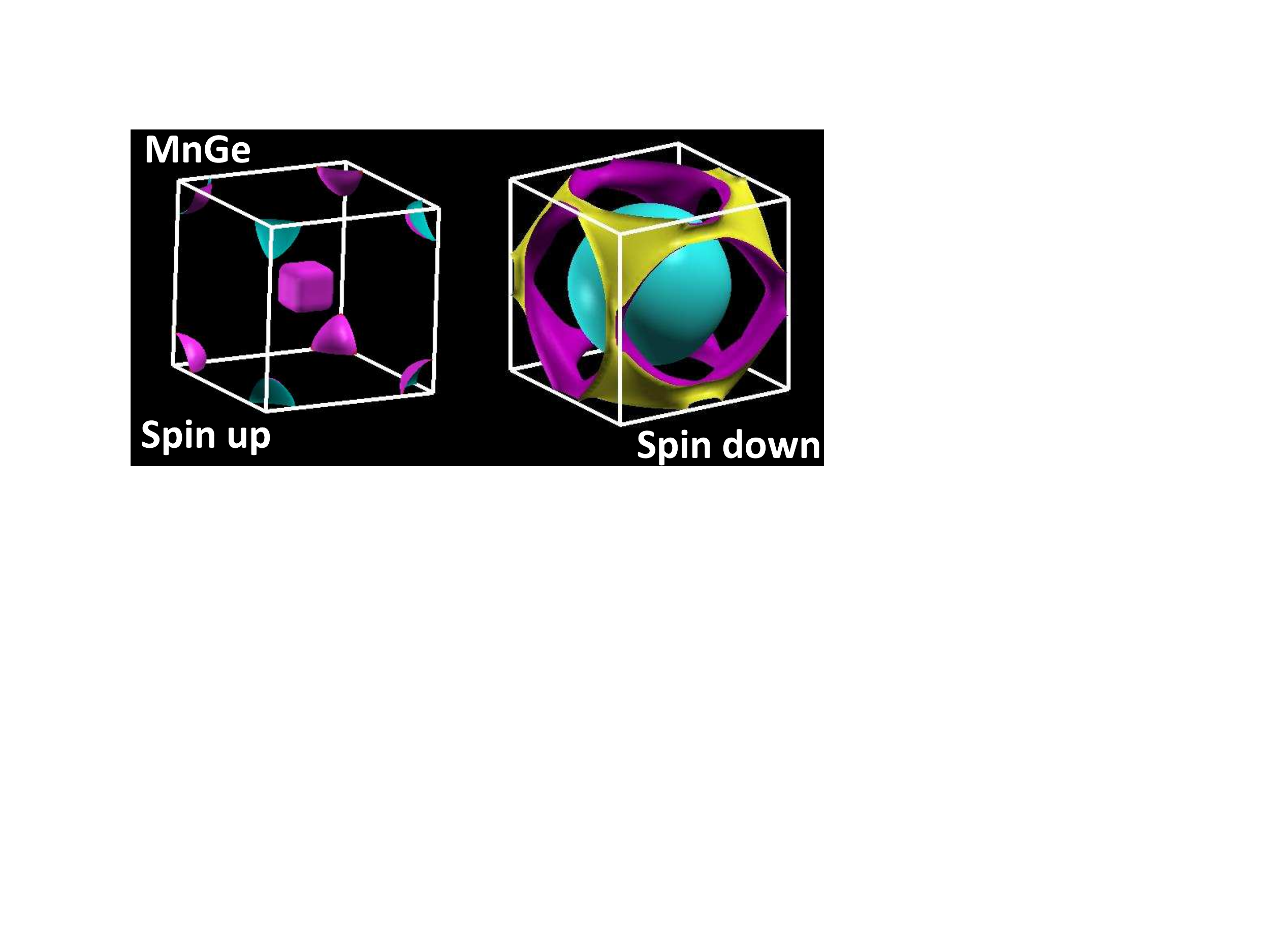}%
  \caption{\label{fig:mngefs} Fermi surface of MnGe. The majority spin
sub-bands (Spin up) are displayed on the left and minority spin
sub-bands (Spin down) are shown on the right. }
\end{figure}

\begin{figure}[htb]
  \includegraphics[angle=0,width=3.2in,bb=45 70 670
  510,clip]{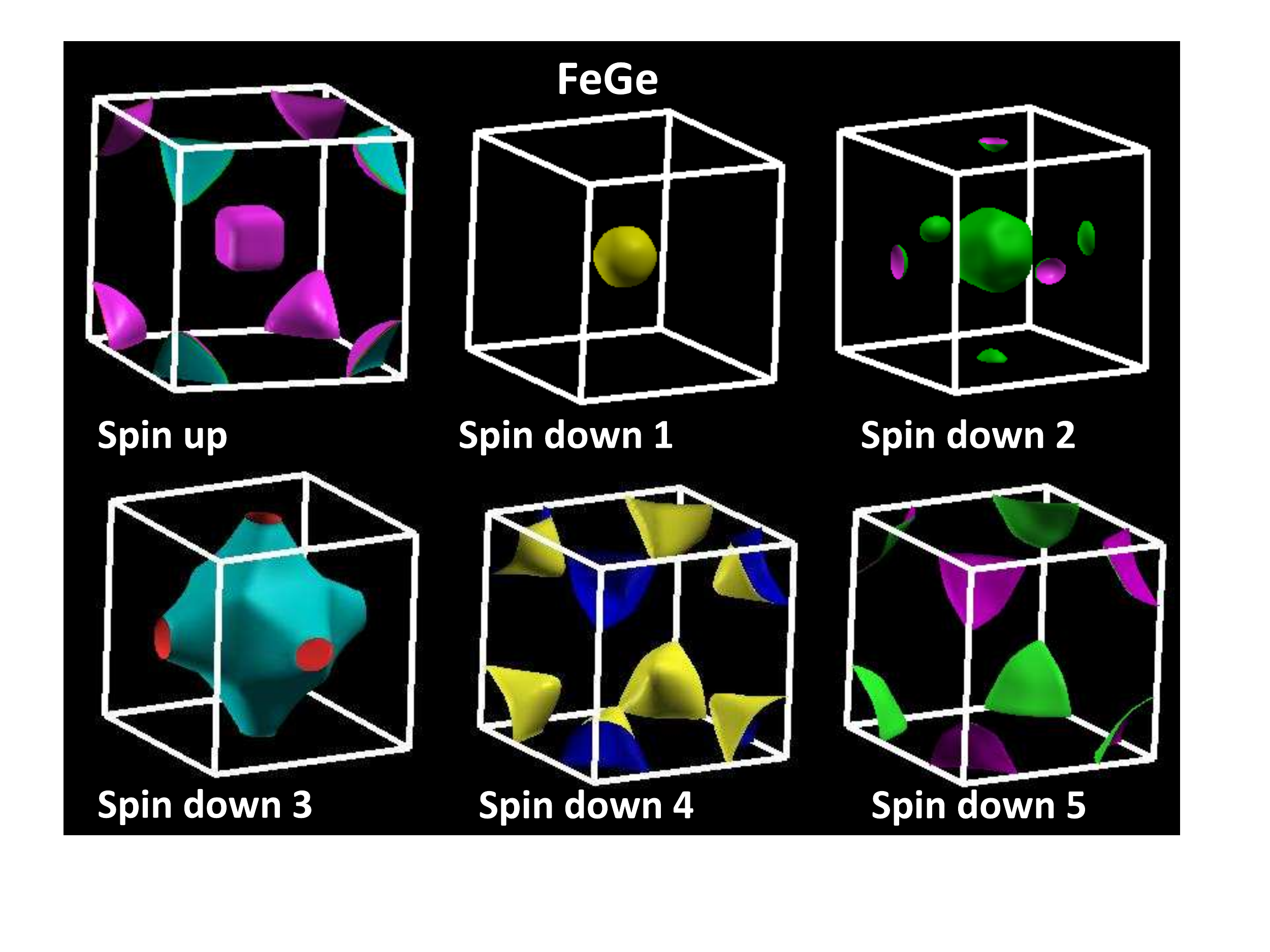}%
  \caption{\label{fig:fegefs} Fermi surface of FeGe. The majority
 spin sub-bands (Spin up) are shown in the top left frame and the five
 minority spin sub-bands (Spin down) are displayed in remaining frames.
 }
\end{figure}

\begin{figure}[htb]
  \includegraphics[angle=0,width=3.2in,bb=185 110 560
  460,clip]{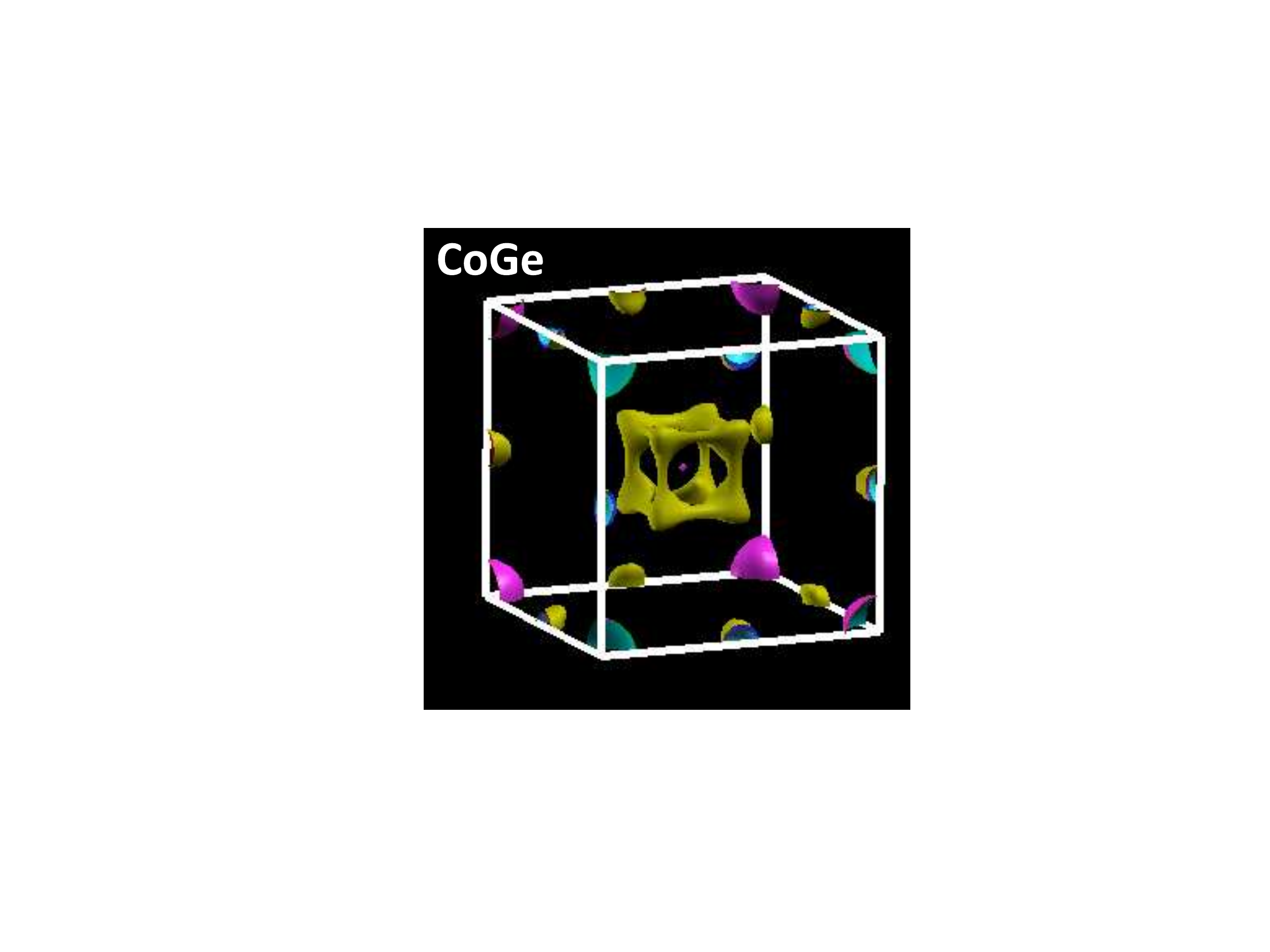}%
  \caption{\label{fig:cogefs} Fermi surface of CoGe. }
\end{figure}

Finally, we present plots of the Fermi surfaces of MnGe in
Fig.~\ref{fig:mngefs}, FeGe in Fig.~\ref{fig:fegefs}, and CoGe in
Fig.~\ref{fig:cogefs}. This comparison highlights both the
similarities in the majority spin sub-bands of MnGe and FeGe with the
electronic structure of CoGe, as pointed out earlier, and the
dominance of the carriers in the spin minority bands of MnGe and
FeGe. This last point is made clear by the small pockets of FS in the
spin up bands of MnGe and FeGe as compared to the respective spin down
bands. The small FS sheets in these sub-bands reflects the tendency
for the spin up FS, along with the single FS in CoGe, to lie near the
same minimum in the DOS demonstrated in Fig.~\ref{fig:bs2}. Thus,
these Fermi surfaces consist of a small pocket of electrons near
$\Gamma$ or, in the case of CoGe, a Dirac point represented by the
small red sphere at the center of Fig.~\ref{fig:cogefs}, and two
pockets of electrons at the $R$-point (corner of the BZ shown in the
figures). These pockets are not very spherical, having a significant
octahedral distortion, which is not surprising given the low site
symmetry in this structure. In the case of CoGe there are also two
small cross-like features of holes centered at the $M$-point and a
cage-like structure centered on $\Gamma$ that is associated with a
band that crosses the Fermi level (holes) by only a few meV which is
below the uncertainty inherent in our calculations. Thus, the
existence of this FS sheet is in question. We have checked this result
by including spin-orbit coupling in our calculation of the electronic
band structure of CoGe to better determine which of the small number
of bands at the Fermi level cross, finding no discernible
differences. This is in contrast to the much larger and, in the case
of FeGe, more numerous, FS sheets in the minority spin-bands of MnGe
and FeGe consistent with the large ordered magnetic moment we measure
for both of these below $T_c$.

There are several aspects to these FS sheets that reflect the low
symmetry of the P2$_1$3 space group including the elliptical, rather
than circular, shape of the intersection of the spin-down FS sheets of
MnGe and FeGe with the BZ boundary. This is a direct consequence of
the lack of four-fold rotational symmetry in the B20 structure.
Notice that the orientation of these ellipses rotate by 90$^o$ on
neighboring faces of the cubic BZ. In addition, as has been pointed
out for the FS of MnSi\cite{jeong}, the FS sheets do not always
intersect the BZ boundary at right angles.  As Jeong points out for
MnSi\cite{jeong}, the periodicity of the FS relies instead on the the
symmetry-required degeneracy of two bands at the BZ boundary so that
the two bands smoothly cross over into one another at the BZ boundary.
Thus, although there are specific features of the FS of these
compounds that reflect the symmetry of the crystal structure, there
are no obvious features which reflect the chirality, a handedness or a
left-right asymmetry, of this crystal structure which is thought to be
responsible for the helimagnetism and the nucleation of Skyrmion
lattices in MnGe\cite{kanazawa2} and FeGe\cite{yu2}.

\section{V. Discussion and Conclusions}
We have synthesized and investigated the properties of cubic B20 MnGe,
and CoGe as well as FeGe to explore the magnetic states of MnGe and to
compare with the other B20 transition metal monosilicides. CoGe is a
small carrier density PM metal that is very similar to CoSi and is
unremarkable in many respects. In contrast, the MnGe data presented
here displays many interesting features, particularly in the magnetic
and thermodynamic properties. We summarize these results here and
compare with the other HM B20 {\it TM} silicides and germanides in
order to highlight how unusual the behavior of MnGe is. Our magnetic
susceptibility data, both $\chi$ and $\chi'$ of our polycrystalline
samples and the single crystal separated from the same sample growth,
indicated a field-independent Curie-Weiss behavior above room
temperature with a Weiss temperature of 270 K and a large fluctuating
moment. This is consistent with the sharp peak in $\chi'(H=0)$ at 275
K indicating the formation of a magnetic state. This peak along with a
small contribution to $\chi''$ between 50 and 275 K are suppressed by
magnetic fields of order 1 kOe. $\chi'$ evolves with field first into
a step like feature at lower temperature, and then to broad maximum
above the field necessary to produce a field polarized state. These
features are similar to that discovered in the region of the $A$-phase
in FeGe and MnSi. In FeGe and MnSi a peak associated with the Curie
temperature evolves into a step wise increase in $\chi'$ at a
temperature that decreases with field, accompanied by a broad peak
just above the zero field $T_c$ and then, at higher $H$, to a simple
broad peak whose maximum moves to higher
temperature\cite{bauer,wilhelm1}. However, these features in MnSi and
FeGe reside within a few K of $T_c$, whereas we observe a very wide
$T$ range where they evolve in MnGe. In addition, a broad peak at 180
K is apparent in MnGe that is suppressed by fields of order 10 kOe
that has no counterpart in either MnSi or FeGe.

In addition, to the features noted above, a sharp peak in $\chi'$
appears in MnGe below 150 K for fields greater than 5 kOe. This sharp
signature moves to significantly lower temperatures with the
application of larger fields and is clearly associated with a dramatic
change in the magnetization curves below 150 K. At these temperatures
a broad region intermediate between the low field HM state and a field
induced FM state emerges. The peaks in $\chi'(T,H)$ are accompanied by
an equally sharp, first-order-like, peak in $C_P(T)$ which originates
at $H=0$ near 160 K where two peaks are apparent. These are in
addition to a seemingly unrelated, field independent, peak at 119 K.

All of the features mentioned in the previous paragraphs are either
absent in MnSi and FeGe or are, perhaps, present in the complex
behavior in close proximity to $T_c$. It is well known that the
behavior of MnSi and FeGe near the Curie point are characterized by a
rich phase diagram which includes a HM low-field state that evolves
with field to a conical state and finally a field induced FM state. In
addition, in a limited region of a few K below $T_c$ over a field
range of 1 to 2 kOe, the $A$-phase, where a Skyrmion lattice has been
shown to arise, produces several features in $\chi'$, $\chi''$, and
$C_P$. For the most part the specific heat of MnSi and FeGe are much
simpler than the complex behavior we find in MnGe, displaying a single
peak near $T_c$ which is broadened by field and having subtle
shoulders that are thought to indicate the so-called intermediate
region just above $T_c$. This region is thus far poorly understood and
has been suggested to host a glassy Skyrmion
phase\cite{neubauer1}. Thus, our data indicate that MnGe may be either
more complex than the better known and studied MnSi and FeGe, or that
the $A$-phase and perhaps the glassy Skyrmion phase may exist in MnGe
that are somehow more apparent and cover a much wider field and
temperature range.

The small angle neutron scattering investigation of Kanazawa {\it et
al.}\cite{kanazawa} have already provided an indication that the
latter may be true. These data on polycrystalline samples grown in a
manner similar to our own indicate a transition to a HM like state
below 170 K with a HM wave vector that is larger than either FeGe or
MnSi and that increases with decreased $T$. The changes that they
observe with field and temperature lead the authors to conclude that
Skyrmion lattice state not only exists over a much wider $T$ and $H$
than in MnSi and FeGe, but that it may be the ground state of the
system, that is, existing at zero temperature and field. However,
these impressive data do not indicate a cause for the phase
transitions we observe in the specific heat nor the subtle magnetic
state indicated in our $\chi'$ data between 165 and 275 K. The
realization that the Skyrmion lattice is much more stable in MnGe
suggests that the phase transitions that we observe in $\chi'$ and
$C_P(T,H)$ may very well be related to a symmetry change either within
the Skyrmion lattice state, or with the collapse of the Skyrmion state
as the system transitions to a more standard magnetic state such as a
conical or a field induced FM state.

\begin{figure}[htb]
  \includegraphics[angle=90,width=3.2in,bb=0 0 450
  650,clip]{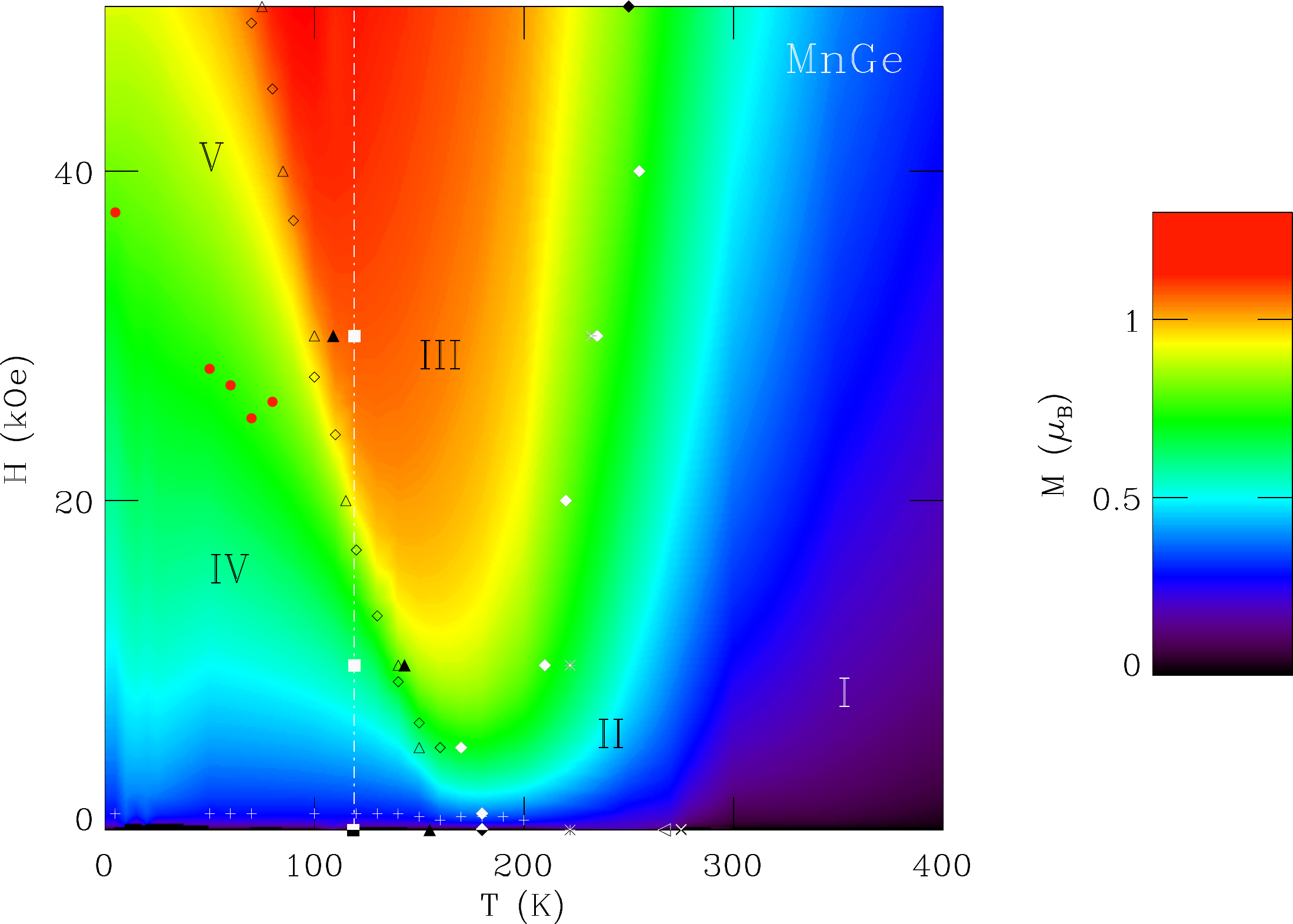}%
  \caption{\label{fig:cont2} Magnetic Phase Diagram.  Contour plot of
the magnetization, $M$, vs.\ temperature, $T$, and field, $H$. Symbols
represent phase transitions or crossover behavior indicated in the
data. Features observed in the specific heat ($C$) in
Fig.~\protect{\ref{fig:sphfig2}} included are the $H$-independent phase
transition (filled squares), the $H$-dependent phase transition
(black filled triangles), the broad maximum observed near 225 K (asterisks),
and the peak near the Curie $T$ (left pointing triangle). Features
apparent in the $T$-dependence of the ac susceptibility, $\chi'$,
Fig.~\protect{\ref{fig:chifig}} include the Curie $T$ (x),
the sharp peaks apparent at finite field at $T< 150$ K (open
triangles), and the broad maxima at $T>150$ K (white and black
diamonds). Maxima apparent in $dM/dH$ determined from the data in
Fig.~\protect{\ref{fig:magfig}} are shown as open diamonds and red
bullets. Also observed in Fig.~\protect{\ref{fig:magfig}} is a region
of large $dM/dH$ at low fields at low $H$s (+). Roman numerals denote 
regions between these features for identification purposes.}
\end{figure}

We summarize what we have learned about the magnetic phase diagram of
MnGe in Fig.~\ref{fig:cont2}(a) and (b) where we plot the
magnetization in a contour plot. We have included benchmarks for the
transitions we observe with with symbols denoting maxima in $\chi'$
and $dM/dH$ as well as the peaks identified in $C_P$.  The phases
denoted include region $I$, the high $T$ PM phase from which we have
deduced a fluctuating moment of 1.35 $\mu_B/$FU and a Weiss $T$ of 275
K. Region $II$ is the ill-defined magnetic state that occurs below
$T_c=275$ K that displays a transition to a field induced FM state at
higher fields (above 15 kOe at 200 K, region $III$ in the figure).  We
make this assignment based upon the sharp peak in $\chi'$ at low
fields and the substantial $M(H)$ above 200 K. Region $IV$ denotes the
low-$H$, low $T$ phase that is identified in the neutron scattering
experiments\cite{kanazawa,makarova,kanazawa2} as either a HM phase or
a Skyrmion lattice phase. A first order phase transition as either $T$
or $H$ are increased is required to enter phases $II$ or $III$ from
region $IV$. The field induced FM phase is identified as region
$III$. Region $IV$ appears to have a very similar low-$H$ behavior to
region $II$ despite there being 2 phase transitions (one field
dependent and one field independent) separating these two phases. At
$T<100$ K a broad maximum in $dM/dH$ is observed near 30 kOe which is
indicated by the red bullets. We note that the Skyrmion phase
identified in Ref.~\cite{kanazawa2} becomes disordered in this field
range so that its signature in the small angle neutron scattering data
is not apparent above these fields (region $V$).  The filled squares
indicate the phase transition that we have observed at 119 K.

The low $T$ phases, regions $IV$ and $V$ in Fig.~\ref{fig:cont2}
display substantial differences from the isostructural compounds FeGe
and MnSi. As we pointed out above, a much larger field is required to
for saturate the magnetization in MnGe, consistent with a much
smaller, $T$-dependent, helical wavelength\cite{kanazawa}. MnSi and
FeGe both display a hysteresis much like that shown in
Fig.~\ref{fig:magfig} for initial field sweep after zero field
cooling. In addition, in MnSi and FeGe there is no evidence, that we
are aware of, for a distinct phase transition in specific heat
measurements at temperatures substantially below $T_c$ at any field as
we have observed in MnGe.
 
The transport properties of MnGe are similar in many ways to that of
isostructural FeGe. In each case a metallic $\rho(T)$ with a RRR of
about 10 along with a small negative MR is found for $T$ exceeding
$T_c$ that increases upon cooling to 100 K\cite{yeo,capan}.  This
negative MR has been ascribed to magnetic fluctuation scattering in
MnSi and FeGe and we interpret the MR in MnGe in a similar
manner\cite{kadowaki2}. In addition, at low $T$ a small positive MR is
observed in FeGe and MnGe below 50 K and in MnSi at $T<1$
K\cite{capan,kadowaki2}. The origin of this positive MR has not been
established but has been suggested to be of a semi-classical origin
because the $H$-dependence is consistent with a $H^2$
behavior\cite{kadowaki2}. Below 10 K a $T^2$ dependence of $\rho(T)$
is observed with a magnitude that is consistent with the electronic
contribution to the specific heat (although there are likely
contributions to $C_p$ from magnons in this $T$ range) and is similar
in magnitude to that found in MnSi at ambient
pressure\cite{kadowaki2,capan}.

In conclusion, we have explored the structural, magnetic,
thermodynamic, and transport properties of MnGe and CoGe stabilized by
synthesis at high pressure.  We find in agreement with previous work
that a simple cubic, B20, crystal structure is stabilized that is
common among {\it TM} monosilicides and
monogermanides\cite{wernick}. Although CoGe is a simple low carrier
density metal much like CoSi, MnGe is magnetic with a HM-like
magnetization curve below a $T_c$ of 275 K. It displays interesting
phase transitions at lower temperatures that have not been observed in
the isostructural compounds. In addition, MnGe requires a much larger
field to saturate the magnetization below 150 K than is seen at higher
$T$ or in the other HM B20 silicides and germanides suggesting a
stronger influence of the Dzyaloshinskii-Moriya interaction. The
transport properties of MnGe are much like that observed in FeGe and
MnSi.  Our data point out the need for single crystals and for further
neutron diffraction and small angle neutron scattering experiments to
determine the structure of the magnetic states that we have identified
in MnGe\cite{makarova,kanazawa2}, the least investigated of the {\it
TM} monogermanide and monosilicide family that have yielded so many
compelling discoveries.

JFD acknowledges support from the NSF through DMR1206763.  JYC and BWF
acknowledge the Office of Basic Energy Sciences, US Department of
Energy through the grant DOE-FG02-08ER46528 and NSF-DMR1358975.  PWA
acknowledges support from the DOE through DE-FG02-07ER46420. This
research was supported in part by the World Premier International
Research Center from MEXT; the Grants-in-Aid for Scientific Research
(22246083) from JSPS; and the Funding Program for World-Leading
Innovative R\&D on Science and Technology(FIRST Program) from JSPS.


 \end{document}